\documentclass[11pt,draftcls,onecolumn]{IEEEtran}

\IEEEoverridecommandlockouts
\usepackage[cmex10]{amsmath}
\usepackage{amssymb}
\usepackage{stackrel}

\usepackage{graphicx}
\usepackage{cite}
\usepackage{enumerate}

\newtheorem{defi}{Definition}
\newtheorem{theorem}{Theorem}
\newtheorem{lemma}{Lemma}
\newtheorem{remark}{Remark}

\newcommand{\Xc}{\mathcal{X}}
\newcommand{\Yc}{\mathcal{Y}}

\newcommand{\Sc}{\mathcal{S}}

\newcommand{\Cc}{\mathcal{C}}
\newcommand{\Mc}{\mathcal{M}}
\newcommand{\Tc}{\mathcal{T}}

\newcommand{\Rc}{\mathcal{R}}

\newcommand{\Nc}{\mathcal{N}}

\DeclareMathOperator*{\argmin}{\arg\!\min}

\begin{document}
%
\title{Secrecy Capacity Scaling in\\Large Cooperative Wireless Networks}

%
%

\author{Mahtab~Mirmohseni and~Panagiotis Papadimitratos
\thanks{M. Mirmohseni is with the Department of Electrical Engineering, Sharif University of Technology, Tehran, IRAN (e-mail: mirmohseni@sharif.edu).}
\thanks{P. Papadimitratos is with the Networked Systems Security Group, KTH Royal Institute of Technology, Stockholm, Sweden (e-mail: papadim@kth.se).}
\thanks{The material in this paper has been presented in part in the IEEE INFOCOM, Toronto, Canada, April-May 2014 \cite{MirPapInf14} and in the Iran Workshop on Communication and Information Theory (IWCIT), Tehran, Iran, May 2014 \cite{MirPapIWCIT14}.}}

\maketitle

\begin{abstract}
We investigate \emph{large} wireless networks subject to security constraints. In contrast to point-to-point, interference-limited communications considered in prior works, we propose active cooperative relaying based schemes. We consider a network with $n_l$ legitimate nodes, $n_e$ eavesdroppers, and path loss exponent $\alpha\geq 2$. As long as $n_e^2(\log(n_e))^{\gamma}=o(n_l)$, for some positive $\gamma$, we show one can obtain unbounded secure aggregate rate. This means zero-cost secure communication, given fixed total power constraint for the entire network. We achieve this result through (i) the source using Wyner randomized encoder and a \emph{serial (multi-stage)} block Markov scheme, to cooperate with the relays and (ii) the relays acting as a virtual multi-antenna to apply beamforming against the eavesdroppers. Our simpler \emph{parallel (two-stage)} relaying scheme can achieve the same unbounded secure aggregate rate when $n_e^{\frac{\alpha}{2}+1}(\log(n_e))^{\gamma+\delta(\frac{\alpha}{2}+1)}=o(n_l)$ holds, for some positive $\gamma,\delta$. Finally, we study the improvement (to the detriment of legitimate nodes) the eavesdroppers achieve in terms of the information leakage rate in a large \emph{cooperative} network in case of \emph{collusion}. We show that again the zero-cost secure communication is possible, if $n_e^{(2+\frac{2}{\alpha})}(\log n_e)^{\gamma}=o(n_l)$ holds, for some positive $\gamma$; i.e., in case of collusion slightly fewer eavesdroppers can be tolerated compared to the non-colluding case.
\end{abstract}



\IEEEpeerreviewmaketitle


\section{Introduction}
The open nature of wireless networks makes them vulnerable to eavesdropping attacks; thus, confidentiality is a crucial security requirement. Conventional, cryptographic techniques have drawbacks, e.g., increasing, with the network size, key management complexity. Moreover, they rely on an assumption of limited attacker computational power, while encrypted data may still provide information to attackers (e.g., through traffic analysis). This motivated efforts to complement cryptographic techniques and fueled interest in information-theoretic physical layer security~\cite{ShiChaWuHuaChe11}.

The natural problem is to find the fundamental limits of performance measures, notably the secure rate legitimate nodes can achieve, considering the overhead imposed by satisfying the secrecy constraints. However, even in simple three- or four-node networks, the problem is open \cite{ElgKim11}; the complex nature of large wireless networks with stochastic node distribution in space makes the derivation of exact results intractable. This motivated the investigation of scaling laws, or the asymptotic behavior of the network, to gain useful insights. The problem of finding scaling laws for large wireless networks with $n$ randomly located nodes was first investigated by Gupta and Kumar in \cite{GupKum00}; they showed that multihopping schemes can achieve at most an aggregate rate that scales like $\sqrt{n}$ under an individual (per node) power constraint. Using percolation theory, achievability of linear scaling was shown by Franceschetti \textit{et al}. \cite{FraDouTseThi07}. The main characteristic of this line of works is the assumption of point-to-point communication, where each receiver (not necessarily the final destination) is interested only in decoding the signal of a particular transmitter; all other signals, roughly termed interference, are treated as noise. Therefore, these are mostly referred to as interference-limited channel models. The broadcasting nature of wireless networks makes cooperation easier, though it decreases the security level. Contrary to the interference-limited model, it has been shown that cooperative schemes increase the aggregate rate to a near-linear scaling under individual power constraints and achieve unbounded transport capacity for fixed total power in some cases (in \cite{XieKum04,OzgLevTse07} and follow-up works).

Recently, there is a growing interest in considering how secrecy constraints affect scaling laws of large wireless networks \cite{KoyKokElg12,ZhaFuWan12,VasGoeTow10,CapGoeLiuTow12,SheGoePisTow12}. To best of our knowledge, all these works considered point-to-point interference-limited communications (multi-hopping) \cite{KoyKokElg12,ZhaFuWan12,VasGoeTow10,CapGoeLiuTow12,SheGoePisTow12} to analyze the \emph{secrecy} capacity scaling; no active cooperative or relaying schemes were considered.

In this paper, contrary to the interference-limited models, we allow for arbitrary cooperation among nodes and concentrate on the information-theoretic relaying schemes. With no secrecy constraint, Xie and Kumar in \cite{XieKum04} proposed a strategy of coherent multistage relaying to achieve unbounded transport capacity for fixed total power in low-attenuation networks, i.e., achieving zero energy cost communication. However, when seeking to address secrecy constraints, active cooperation (relaying) is a double-edged sword: it benefits both legitimate receivers and eavesdroppers. Considering this trade-off, the fundamental question is whether zero-cost \emph{secure} communication is possible through active cooperation. We answer this question positively here, filling this theoretical gap. Our result is further motivated by recent technological developments for relaying-based schemes (e.g., massive deployment of relay nodes in LTE-Advanced networks \cite{LTE10,SawKisMorNisTan10}).

\subsection{Background and Related Work}\label{subsec:back}
Physical layer security using information-theoretic tools leverages the channel statistics to thwart eavesdroppers (attackers); depending on the channel conditions, a secure positive rate can be possible if suitable coding schemes are employed. The information-theoretic notion of secrecy was introduced by Shannon in \cite{Sha49}, where he showed that in order to achieve perfect secrecy, i.e., zero information leakage, one needs a secret key of size at least equal to the message size. This result inspired keyless information-theoretic security in a noisy communication model called the wiretap channel. Wyner determined the capacity of the degraded wiretap channel, with the channel to the eavesdropper being a degraded version of the channel to the legitimate receiver \cite{Wyn75}. The wiretap secrecy capacity achieving scheme, known also as \emph{Wyner wiretap channel coding}, comprises multicoding and randomized encoding \cite[Section~22.1.1]{ElgKim11}. Csisz\'{a}r and K\"{o}rner extended the secrecy capacity result to the the general wiretap channel (not necessarily degraded) \cite{CsiKor78}.

There has been considerable recent research interest in multi-user wiretap channels \cite{LiuMarSpaYat08,EkrUlu13,ChiaElG12,Ooh07,LaiElG08,LiaPooYin11,BasUlu12,BasUlu13}. In these channels, cooperation among legitimate users is possible in two different ways. First, through active cooperation: legitimate nodes act as relays and cooperate with the source of the message in transmitting its message to the destination. This scenario with a single relay was introduced in \cite{LaiElG08} as the relay-eavesdropper channel where the secrecy rates were derived using relaying strategies such as the Decode-and-Forward (DF) scheme \cite{CoveElg79}. The case of multiple relays was investigated in \cite{DonHanPetPoo08,BasUlu12}. Second, passive cooperation, also known as deaf cooperation: the so called helper nodes transmit signals, which are independent from the transmitted message of the legitimate transmitter, to confuse the eavesdroppers and increase the secure rates \cite{DonHanPetPoo08,BasUlu13,NegGoe05}. In both cooperation modes, one can try to apply beamforming at the helper nodes to improve the secrecy by constructing the virtual Multiple Input Multiple Output (MIMO) scenarios and/or perform Zero-Forcing (ZF) at eavesdroppers \cite{NegGoe05,LanWor03,shaLiuUlu09,KhiWor10,YukErk11,RezAlo12,OggHas11}. It was shown that in a high-SNR regime, the ZF transmit scheme is Diversity-Multiplexing Tradeoff (DMT) optimal in a MIMO wiretap channel, with three nodes, a source, a destination and an eavesdropper \cite{YukErk11,RezAlo12}. In this paper, we concentrate on the \emph{active cooperation} schemes based on information-theoretic secrecy coding schemes.

The adversarial behavior in the aforementioned scenarios is captured by multiple eavesdroppers that can either listen individually to the channel (\emph{non-colluding} eavesdroppers) or they can share their observations and make the attack more effective (\emph{colluding} eavesdroppers) \cite{PinBarWin09}. The distinction of the two adversarial models is significant. Collusion implies increased sophistication, thus more powerful adversaries. In practice, it may be feasible for many systems. Thus, a non-colluding eavesdroppers model may underestimate the adversary in some applications. In any case, it is an important question: \emph{How does the increase in adversarial power (collusion) affect the secrecy rates and their scaling?} The mitigation of colluding eavesdroppers was investigated \cite{PinBarWin09,KoyKokElg12,ZhaFuWan12,PinBarWin12II}.

Although there is considerable effort in these works on small networks (for both non-colluding and colluding eavesdroppers), consisting of few nodes with deterministic locations, the problem of secure communication in large networks received relatively less attention. Scaling laws for the secure aggregate rate were derived for large wireless networks, only under the assumption of \emph{interference-limited channel}. Koyluoglu \emph{et al.} \cite{KoyKokElg12} recently achieved a secure aggregate rate of scaling $\sqrt{n}$ for a dense network of $n$ legitimate nodes, as long as the ratio of the densities of eavesdroppers and legitimate nodes scales as $(\log{n})^{-2}$, for non-colluding eavesdroppers. While for \emph{colluding} eavesdroppers, the same rate scaling (i.e., $\sqrt{n}$) is achieved for a lower density of eavesdroppers \cite{KoyKokElg12}. The authors in \cite{VasGoeTow10,CapGoeLiuTow12,SheGoePisTow12} considered extended networks with unknown eavesdropper locations and achieved a secure rate of order $1$. This result is achieved through a deaf (passive) cooperative multi-hopping scheme in \cite{SheGoePisTow12}. These scaling results were achieved assuming that the transmission power for each node is fixed. Thus, the total power scales linearly with the number of nodes, $n$, and the cost of secure communication (defined as the total power over the secure rate) goes to $\infty$.

\subsection{Our Contributions}\label{subsec:contributions}
Our work is the first that allows \emph{arbitrary} cooperation among legitimate nodes in deriving scaling laws for large wireless networks with secrecy constraints. Without the limitation of point-to-point communication, we show that cooperation can achieve \textit{unbounded secure rate with fixed total power}, i.e., \textit{zero-cost secure communication}, as long as the number of the eavesdroppers is less than a \emph{derived} threshold. We consider a \emph{dense} network, with a static path loss physical layer model and path loss exponent $\alpha\geq 2$, and stochastic node placement. $n_l$ legitimate nodes and $n_e$ eavesdroppers are distributed according to Poisson Point Processes (PPP) with intensities $\lambda_l$ and $\lambda_e$, respectively, in a square of unit area. We consider the fixed total power constraint and find two scaling results for $\frac{n_e}{n_l}$, where by satisfying these results one can obtain infinite secure aggregate rate and thus zero-cost secure communication. Compared to \cite{XieKum04}, this means that $n_e$ eavesdroppers can be tolerated asymptotically and do not affect the communication cost.

To achieve this result, we make use of (i) block Markov DF relaying, (ii) Wyner wiretap coding at the source to secure the new part of the message transmitted in each block, and (iii) beamforming, to secure the coherent parts transmitted cooperatively by all the nodes in the network. To apply DF, we propose two types of schemes: parallel (two-stage) relaying and serial (multi-stage) relaying. For beamforming, partial ZF at the eavesdroppers is used. DF based strategies for multiple relay networks were proposed in \cite{XieKum04,XieKum05} and then they were extended to such networks with an eavesdropper in \cite{BasUlu12}, where some ZF schemes were applied. Here, we first extend these schemes to our network model with stochastic distribution of legitimate nodes and eavesdroppers by deriving the conditions under which we can apply the schemes. The main challenges we face are: the selection of relays among the legitimate nodes, the priority and power allocation, and the choice of appropriate beamforming parameters. Once these challenges addressed, we utilize the derived rates to achieve zero-cost secure communication.

Using the parallel (two-stage) relaying strategy, we show the possibility of achieving unbounded secure aggregate rate as long as $n_e^{\frac{\alpha}{2}+1}(\log(n_e))^{\gamma+\delta(\frac{\alpha}{2}+1)}=o(n_l)$ for some positive $\gamma,\delta$ holds. Our scheme has two stages. First, the source of the message transmits to $n_r$ relay nodes within some distance. At the second stage, the source and these relay nodes use block Markov coding \cite{ElgKim11} to cooperatively transmit the message to the destination, while using ZF against the eavesdroppers. In fact, relay nodes can be seen as a distributed virtual multi-antenna; using this diversity to combat the eavesdroppers. Transmissions are pipelined and relay nodes operate in a full-duplex mode, a typical assumption (e.g., \cite{KraGasGup05,XieKum04}).

At the expense of additional complexity, we tolerate even more eavesdroppers with serial (multi-stage) relaying. We achieve zero energy cost secure communication as long as $n_e^2(\log(n_e))^{\gamma}=o(n_l)$ holds, for some $\gamma>0$. In this scheme, all network nodes can act as relays for the source node but they are ordered in clusters and use block Markov coding and coherent transmission. Nodes in each cluster form a virtual multi-antenna to apply ZF at the eavesdroppers.

Finally, we investigate how a more powerful adversary model (i.e., \emph{colluding} eavesdroppers) degrades the scaling of $\frac{n_e}{n_l}$ for cooperative networks.
We show that, even in the presence of \emph{colluding} eavesdroppers, active cooperation achieves zero-cost secure communication while tolerating less eavesdroppers (compared to the non-colluding case). We let eavesdroppers exchange their channel outputs (observations), i.e., collude, for free; this is the perfect collusion model considered in literature \cite{PinBarWin09,KoyKokElg12}. We achieve an unbounded secure rate given fixed total power (for the entire network), as long as $n_e^{(2+\frac{2}{\alpha})}(\log n_e)^{\gamma}=o(n_l)$ holds for some $\gamma>0$. For the achievability, we  propose a \emph{serial (multi-stage)} relaying based scheme.

The rest of the paper is organized as follows. Section~\ref{sec:definition} introduces the network model and notation. Section~\ref{sec:Par} describes our proposed parallel relaying scheme and its scaling is derived. In Section~\ref{sec:Ser}, the results of serial relaying scheme are stated. A number of remarks are provided in Section~\ref{sec:discussion}.

\section{Network Model and Preliminaries}\label{sec:definition}
\textbf{Notation}: Upper-case letters (e.g., $X$) denote Random Variables (RVs) and lower-case letters (e.g., $x$) their realizations. The probability mass function (p.m.f) of a RV $X$ with alphabet set $\Xc$ is denoted by $p_X(x)$; occasionally subscript $X$ is omitted.
$A_\epsilon^n(X,Y)$ is the set of $\epsilon$-strongly, jointly typical sequences of length $n$.
$X^j_i$ indicates a sequence of RVs $(X_i,X_{i+1},...,X_j)$; we use $X^j$ instead of $X^j_1$ for brevity. $\Cc\Nc(0,\sigma^2)$ denotes a zero-mean complex value Gaussian distribution with variance $\sigma^2$. The variables related to the legitimate nodes and eavesdroppers are indicated with sub/superscripts $l$ and $e$, respectively. $\|\mathbf{X}\|_p$ is the $L^p$-norm of a vector $\mathbf{X}$; $\mathbf{X}(i)$ is its $i$th element. $(\cdot)^T$, $(\cdot)^\dag$ and $\Nc(\cdot)$ denote the transpose, conjugate transpose and null space operations, respectively.
For stating asymptotic results (Landau notation), $f(n)=o(g(n))$ if $\lim\limits_{n\rightarrow \infty}\frac{f(n)}{g(n)}\rightarrow 0$.

\textbf{Network and adversary model}:
We consider a dense wireless network, with channel gains obeying a static path loss model, decaying exponentially as the distance between the (stochastically distributed) nodes increases. This  is consistent with models in prior works on capacity scaling laws \cite{GupKum00,FraDouTseThi07,XieKum04,OzgLevTse07} and secrecy capacity scaling \cite{KoyKokElg12}.
For the adversary model, we consider two cases: \emph{non-colluding} and \emph{perfect colluding} passive eavesdroppers (as per all existing large network analyses modeling collusion \cite{KoyKokElg12,PinBarWin09,PinBarWin12II}).
For brevity, in the rest of the paper, the eavesdroppers are non-colluding, unless it is stated otherwise explicitly.

The network is a square of unit area where both the legitimate nodes and \textit{eavesdroppers} are placed, according to Poisson Point Processes (PPP) with intensities $\lambda_l$ and $\lambda_e$, respectively, which is a suitable assumption when nodes are independently and uniformly distributed in the network area or there is a substantial mobility \cite{WebAndJin10}.
There is a set $\Nc_l$ of legitimate nodes and their number is $n_l=|\Nc_l|$. Similarly, $\Nc_e$ and $n_e=|\Nc_e|$ are the set of eavesdroppers and their number. As we consider large-scale networks, throughout the paper, we implicitly assume that $n_l$ and $n_e$ go to $\infty$.
Each legitimate node $i\in\Nc_l$ can be a source of message $m_i\in\Mc_i=[1:2^{n_tR_i}]$ and send it to its randomly chosen destination $j\in\Nc_l\setminus\{i\}$ in $n_t$ channel uses. Every legitimate node $i\in\Nc_l$ operates in a full-duplex mode; at time slot $t$, it transmits $X_i(t)$ and receives $Y_i^l(t)$. The set of transmitting nodes at time slot $t$ is denoted by $\Tc(t)\subseteq\Nc_l$. As we consider passive attackers, each eavesdropper $j\in\Nc_e$ only observes the channel and at time slot $t$, it receives $Y_{j}^e(t)$. Therefore,
\begin{IEEEeqnarray}{rcl}
    Y_{i}^l(t)&=&\sum\limits_{k\in\Tc(t)\setminus\{i\}}{h_{k,i}^l(t)X_k(t)}+Z_{i}^l(t)\label{eqn:RxLeg}\\
    Y_{j}^e(t)&=&\sum\limits_{k\in\Tc(t)}{h_{k,j}^e(t)X_k(t)}+Z_{j}^e(t)\label{eqn:RxEav}
\end{IEEEeqnarray}
where, for any $i\in\Nc_l\setminus\{k\}$ and $j\in\Nc_e$, the static path loss model channel gains are given by:
\begin{IEEEeqnarray}{rcl}\label{eqn:chgain}
    h_{k,i}^l(t)=(d_{k,i}^l)^{-\alpha/2} &\quad,\quad& h_{k,j}^e(t)=(d_{k,j}^e)^{-\alpha/2}
\end{IEEEeqnarray}
with $d_{k,i}^l$ and $d_{k,j}^e$ denoting the distances between the transmitter $X_k,k\in\Tc(t)$ and the receiver $Y_i^l$ and $Y_j^e$, respectively. $X_k(t),k\in\Tc(t),t\in[1:n_t]$ is an input signal and we consider the total power constraint in the network:
\begin{IEEEeqnarray}{rcl}\label{eqn:tot_power_cons}
    \frac{1}{n_t}\sum\limits_{t=1}^{n_t}\sum\limits_{k\in\Tc(t)}|x_{k}(t)|^2\leq \overline{P}_{tot}.
\end{IEEEeqnarray}
Moreover, $Z_{i}^l(t)$ and $Z_{i}^e(t)$ are independent and identically distributed (i.i.d) and zero mean circularly symmetric complex Gaussian noise components with powers $N^l$ and $N^e$, i.e., $Z_{i}^l\sim\Cc\Nc(0,N^l)$ and $Z_{i}^e\sim\Cc\Nc(0,N^e)$, respectively. Our \textbf{network model}, defined above, is called \emph{Secure Network} ($\Sc\Nc$) throughout the paper.

To model collusion, in addition to observing the channel ($Y_{j}^e(t)$ for $j\in\Nc_e$ at time slot $t$), the eavesdroppers can exchange their observations for free (because of the perfect collusion assumption). This means that all eavesdroppers have access to all the observations, shown by the vector $\mathbf{Y}^e(t)$, with $Y_{j}^e(t)$ its $j$-th element. We term our network model in this case, \emph{$\Sc\Nc$ with Perfect Colluding Eavesdroppers ($\Sc\Nc$-PCE)} throughout the rest of the paper.

\begin{defi}\label{def:code}
Let $\mathbf{R}=[R_i:i\in\Nc_l]$ be the rate vector and $2^{n_t\mathbf{R}}\doteq\{2^{n_tR_i}:i\in\Nc_l\}$. A $(2^{n_t\mathbf{R}},n_t,P_e^{(n_t)})$ code for $\Sc\Nc$ (or $\Sc\Nc$-PCE) consists of
\begin{enumerate}[(i)]
  \item $n_l$ message sets $\Mc_i=[1:2^{n_tR_i}]$ for $i\in\Nc_l$, where $m_i$ is uniformly distributed over $\Mc_i$.
  \item $|\Tc(t)|$ sets of \emph{randomized} encoding functions at the transmitters: $\{f_{i,t}\}_{t=1}^{n_t}:\mathbb{C}^{t-1}\times\Mc_i\longrightarrow \mathbb{C}$ such that $x_{i,t}=f_{i,t}(m_i,y_{l_i}^{t-1})$, for $i\in\Tc(t)$, $1\leq t\leq n_t$ and $m_i\in\Mc_i$.
  \item Decoding functions, one at each legitimate node $i\in\Nc_l$, $g_{i}: (\Yc_{i}^l)^{n_t}\times\Mc_{i}\mapsto\Mc_{k}$ for some $k\in\Nc_l\setminus\{i\}$, where it is assumed that node $i$ is the destination for the message of source $k$.
  \item Probability of error for this code is defined as $P_e^{(n_t)}=\max\limits_{i\in\Nc_l}P_{e,i}^{(n_t)}$ with:
        \begin{align}\label{eqn:def_Pe}
             \!\!\!\!\!\!P_{e,i}^{(n_t)}=\frac{1}{2^{n_t\|\mathbf{R}\|_1}}\sum\limits_{m_k\in\mathfrak{M}}{Pr(g_{i}((Y_{i}^l)^{n_t},m_{i})\neq m_k | \mathfrak{M}\textrm{ sent})}
        \end{align}
        where $\mathfrak{M}=\{m_i:i\in\Nc_l\}$.
  \item For $\Sc\Nc$: The information leakage rate for eavesdropper $j\in\Nc_e$ is defined as
\begin{align}\label{eqn:def_leakage}
    R_{L,j}^{(n_t)}=\frac{1}{n_t}I(\mathfrak{M};(Y_j^e)^{n_t}).
\end{align}

For $\Sc\Nc$-PCE: The information leakage rate for the perfect colluding eavesdroppers set $\Nc_e$ is defined as
\begin{align}\label{eqn:def_leakage_col}
    R_{L}^{(n_t)}=\frac{1}{n_t}I(\mathfrak{M};(\mathbf{Y}^e)^{n_t}).
\end{align}
\end{enumerate}
\end{defi}

\begin{defi}\label{def:rate}
For $\Sc\Nc$: A rate-leakage vector $(\mathbf{R},\mathbf{R_L})$ is achievable if there exists a sequence of $(2^{n_t\mathbf{R}},n_t,P_e^{(n_t)})$ codes such that $P_e^{(n_t)}\rightarrow 0$ as $n_t\rightarrow\infty$ and $\limsup\limits_{n_t\rightarrow\infty} R_{L,j}^{(n_t)}\leq \mathbf{R_L}(j)$. The secrecy capacity region, $\Cc_s$, is the region which includes all achievable rate vectors, $\mathbf{R}$, such that perfect secrecy is achieved, i.e., $\mathbf{R_L}=\mathbf{0}$.

For $\Sc\Nc$-PCE: A rate vector-leakage pair $(\mathbf{R},R_L)$ is achievable if there exists a sequence of $(2^{n_t\mathbf{R}},n_t,P_e^{(n_t)})$ codes such that $P_e^{(n_t)}\rightarrow 0$ as $n_t\rightarrow\infty$ and $\limsup\limits_{n_t\rightarrow\infty} R_{L}^{(n_t)}\leq R_L$. The secrecy capacity region $\Cc_s$ includes all achievable rate vectors, $\mathbf{R}$, such that perfect secrecy is achieved, i.e., $R_L=0$.

In large-scale networks, it is intractable to consider the $n_l$-dimensional secrecy capacity region; thus, we focus on the secure aggregate rate, defined as:
\begin{align}\label{eqn:def_agg_rate}
\Rc_{s}=\sup\limits_{\mathbf{R}\in\Cc_s}\|\mathbf{R}\|_1.
\end{align}
\end{defi}
As we are interested in the achievability of $\Rc_{s}$, without loss of generality, we assume that only one source-destination pair is active and the other nodes assist their transmission. Therefore, we set $|\mathfrak{M}|=1$. Without loss of generality, we assume that node 1 is the source node, i.e., $\mathfrak{M}=\{m_1\}$: it transmits $X_1(t)$; and $Y_{1}^l(t)=\emptyset$. Thus, $\Rc_{s}=R_1$. We denote the destination of $m_1$ by $n_l$-th node: it receives $Y_{n_l}^l(t)$; and $X_{n_l}(t)=\emptyset$. This means that the transmitter $X_1$ wishes to send message $m_1\in\Mc_1=[1:2^{n_tR_1}]$ to the receiver $Y_{n_l}^l$ with the help of nodes in $\Nc_l\setminus\{1,n_l\}$, while keeping it secret from the eavesdroppers in $\Nc_e$.
\begin{remark}\label{remark:agg_rate}
If a secure aggregate rate $\Rc_{s}=R_1$ is achievable in the above scenario (with uniformly random matching of the source-destination pairs), any rate vector $\mathbf{R}$ with $\|\mathbf{R}\|_1=\Rc_{s}$ is also achievable using a time-sharing scheme. For example, consider a network of $n$ nodes with a total rate of 1 bit/sec. If there is only one active source-destination pair, the source can transmit at the rate of 1 bit/sec. Otherwise, a Time Division Multiple Access (TDMA) scheme, with $n$ equal time slots, achieves the rate of $\frac{1}{n}$ for each (source) node in the network, with the total rate of $n\times\frac{1}{n}=1$ bit/sec. Any other rate allocation with unit total rate is also attainable by using TDMA with non-equal time slots.
\end{remark}

\section{Parallel Relaying}\label{sec:Par}
In this section, we consider a parallel (two-stage) relaying scheme and obtain the maximum number of eavesdroppers that can be tolerated in a zero-cost secure communication. In fact, our main result of this section, Theorem~\ref{thm:SL_par} shows that we achieve an unbounded secure aggregate rate for a fixed total power as long as $n_e^{\frac{\alpha}{2}+1}(\log(n_e))^{\gamma+\delta(\frac{\alpha}{2}+1)}=o(n_l)$, for some positive $\gamma,\delta$. Our proof is derived in three steps:
\begin{enumerate}
  \item First, we provide a lower bound on the secrecy capacity achieved through active cooperation, randomized encoding and beamforming in Theorem~\ref{thm:Ach_par_DF}. We propose a two-stage DF relaying and design the appropriate codebook mapping that enables ZF at the eavesdroppers. To apply these strategies, we derive conditions on the number and location of the relay nodes.
  \item In the second step, the main challenge is to find strategies to apply the achievability scheme of the first step to our network model ($\Sc\Nc$). In Lemma~\ref{lemma:par_squ_nodes}, we obtain the constraints on the number of legitimate nodes and eavesdroppers under which our network satisfies the conditions of the first step and the achievability scheme can be applied.
  \item In the last step, we apply the fixed total power constraint and show that the achievable secure aggregate rate of the first step can be unbounded and derive the maximum number of the eavesdroppers which can be tolerated in Theorem~\ref{thm:SL_par}.
\end{enumerate}
\textbf{Step 1: }As mentioned in Section~\ref{sec:definition}, the achievability relies on a single unicast scenario. Recall that a lower bound on the secrecy capacity of this scenario is an achievable secure aggregate rate for $\Sc\Nc$. Here, $n_r$ relay nodes (in $\Nc_l\setminus\{1,n_l\}$) are used as specified in the following theorem.
\begin{theorem}\label{thm:Ach_par_DF}
For $\Sc\Nc$, if there exists a set of transmitters
\begin{IEEEeqnarray}{c}
\Tc=\Big\{1,\big\{i\:\Big|\:|h_{1,i}^l|^2\geq \max\{\frac{N^l}{N^e}|h_{1,j}^e|^2,|h_{1,{n_l}}^l|^2\}\big\}\Big\}\label{eqn:def_Tx_par}
\end{IEEEeqnarray}
such that $n_r=|\Tc|-1 \geq n_e$, the following secure aggregate rate is achievable :
\begin{IEEEeqnarray}{rl}
\Rc_s^{DF,ZF,par}=\max_{\mathbf{B},\tilde{P}_1,\tilde{P}_u}\min\limits_{j\in\Nc_e}\min\{&\log(\frac{N^e}{N^l}\frac{N^l+|h_{1,i^*}^l|^2\tilde{P}_1}{N^e+|h_{1,j}^e|^2\tilde{P}_1}),\nonumber\\
&\log(\frac{N^e}{N^l}\frac{N^l+|h_{1,{n_l}}^l|^2\tilde{P}_1+|\sum\limits_{k\in\Tc}h_{k,{n_l}}^l\beta_k|^2\tilde{P}_u}{N^e+|h_{1,j}^e|^2\tilde{P}_1})\}\label{eqn:Ach_par_ZF}
\end{IEEEeqnarray}
where
\begin{IEEEeqnarray}{lll}
i^*=\argmin\limits_{i\in\Tc\setminus\{1\}}|h_{1,i}|&&\label{eqn:ach_ZF_cond_rel}\\
\beta_k=\mathbf{B}(k)&\textrm{ where}\quad&\mathbf{B}\in {\mathcal{N}(\mathbf{H}_{\Nc_e,\Tc})}\label{eqn:ach_ZF_cond_coef}\\
\tilde{P}_1+\|\mathbf{B}\|_2^2\tilde{P}_u\leq \overline{P}_{tot}&&\label{eqn:ach_ZF_cond_pow}
\end{IEEEeqnarray}
in which $\mathbf{H}_{\Nc_e,\Tc}\in\mathbb{C}^{n_e\times (n_r+1)}$ is the transmitters-eavesdroppers channel matrix whose $(j,i)$-th element is $h_{i,j}^e$ for $i\in\Tc,j\in\Nc_e$.
\end{theorem}
\begin{IEEEproof}
First, we outline the coding strategy, based on a two-stage block Markov coding, i.e., all relays have the same priority for the source. In each block, the source sends the fresh message to \emph{all} $n_r$ relay nodes and uses Wyner wiretap coding to keep this part of the message secret from the eavesdroppers. At the same time, the source and the relays cooperate in sending the message of the previous block by coherently transmitting the related codeword. This coherent transmission enables them to use ZF against the eavesdroppers, by properly designed beamforming coefficients. As the cooperative codewords of the relays are fully zero-forced at all eavesdroppers, no Wyner wiretap coding is needed at the relays.

Now, to apply this coding strategy, first we provide achievable rate $\Rc_s^{DM,par}$ based on two-stage block Markov coding (parallel DF relaying) and Wyner wiretap coding for the general discrete memoryless channel in Lemma~\ref{lemma:Ach_par_DF_DM} (proof is provided in Appendix~\ref{app:proof_lemma:Ach_par_DF_DM}). Then, we extend $\Rc_s^{DM,par}$ to the Gaussian channel in Lemma~\ref{lemma:Ach_par_DF} and derive $\Rc_s^{DF,par}$ (proof in Appendix~\ref{app:proof_lemma:Ach_par_DF}). Finally, we apply ZF on $\Rc_s^{DF,par}$ to achieve the desired result, i.e., $\Rc_s^{DF,ZF,par}$. For simplicity in notation, let $\Nc_l=\{1,\ldots,n_l\}$, $\Tc=\{1,\ldots,n_r+1\}$ and $\Nc_e=\{1,\ldots,n_e\}$.
\begin{lemma}\label{lemma:Ach_par_DF_DM}
For the general discrete memoryless counterpart of $\Sc\Nc$, given by some conditional distribution $p(y_2^l,\ldots,y_{n_l}^l,y_{1}^e\ldots,y_{{n_e}}^e|x_1,\ldots,x_{n_l})$, the secrecy capacity is lower-bounded by:
\begin{IEEEeqnarray}{l}\label{eqn:Ach_par_DF_DM}
\Rc_s^{DM,par}=\sup\min\limits_{j\in\Nc_e}\{\min\{\min\limits_{i\in\Tc\setminus \{1\}}\!\!I(U_1;Y_{i}^l|U), I(U,U_1;Y_{n_{l}}^l)\}-I(U,U_1;Y_{j}^e)\}
\end{IEEEeqnarray}
where the supremum is taken over all joint p.m.fs of the form
\begin{IEEEeqnarray}{c}\label{eqn:Ach_par_DF_DM_pmf}
p(u,u_1)p(x_1,\ldots,x_{n_r+1}|u,u_1).
\end{IEEEeqnarray}
\end{lemma}

Now, we extend the above lemma to accommodate our model ($\Sc\Nc$). Even for a simple channel with one relay and one eavesdropper, the optimal selection of the RVs in Lemma~\ref{lemma:Ach_par_DF_DM} (i.e., finding the optimal p.m.f of \eqref{eqn:Ach_par_DF_DM_pmf}) is an open problem \cite{BasUlu12}. Hence, we propose an appropriate suboptimal choice of input distribution, using Gaussian RVs, to achieve the following rate.
\begin{lemma}\label{lemma:Ach_par_DF}
The following secure aggregate rate is achievable for $\Sc\Nc$:
\begin{IEEEeqnarray*}{rl}
\Rc_s^{DF,par}=\max_{\mathbf{B},\tilde{P}_1,\tilde{P}_u}\min\limits_{j\in\Nc_e}\big\{&\min\{\min\limits_{i\in[2:n_r+1]} \log(1+\frac{|h_{1,i}^l|^2\tilde{P}_1}{N^l}),\log(1+\frac{|h_{1,{n_l}}^l|^2\tilde{P}_1+|\sum\limits_{k=1}^{n_r+1}h_{k,{n_l}}^l\beta_k|^2\tilde{P}_u}{N^l})\}\\
&-\log(1+\frac{|h_{1,j}^e|^2\tilde{P}_1+|\sum\limits_{k=1}^{n_r+1}h_{k,j}^e\beta_k|^2\tilde{P}_u}{N^e})\big\}\yesnumber\label{eqn:Ach_par_DF}
\end{IEEEeqnarray*}
where $\beta_k=\mathbf{B}(k)$ and $\tilde{P}_1+\|\mathbf{B}\|_2^2\tilde{P}_u\leq \overline{P}_{tot}$.
\end{lemma}

It can easily be seen from \eqref{eqn:Ach_par_DF} that to have a positive secrecy rate the source-relay links should be stronger than the source-eavesdropper links. Moreover, for the DF strategy to be better than point-to-point transmission the source-relay links should be stronger than the direct source-destination link. Therefore, these two conditions ``select'' the $n_r$ relay nodes in the DF strategy and, hence, the set of transmitters $\Tc(t)$ given by \eqref{eqn:def_Tx_par}. Moreover, the condition in \eqref{eqn:ach_ZF_cond_rel} is obtained by considering the inner $\min$ in \eqref{eqn:Ach_par_DF}.

Returning to \eqref{eqn:Ach_par_DF}, one should determine the beamforming coefficient vector $\mathbf{B}$. Finding the closed form solution is an open problem \cite{DonHanPetPoo08}. Thus, we consider a suboptimal strategy by applying ZF at all eavesdroppers and obtain
\begin{IEEEeqnarray}{c}
\mathbf{H}_{\Nc_e,\Tc}\mathbf{B}=\mathbf{0}\label{eqn:ZF_mat}
\end{IEEEeqnarray}
where $\mathbf{H}_{\Nc_e,\Tc}$ is defined in Theorem~\ref{thm:Ach_par_DF}. Hence, the coefficient vector $\mathbf{B}$ must lie in the null space of $\mathbf{H}_{\Nc_e,\Tc}$, as stated in \eqref{eqn:ach_ZF_cond_coef}. By applying \eqref{eqn:ZF_mat} to \eqref{eqn:Ach_par_DF}, we achieve \eqref{eqn:Ach_par_ZF}.

In order to ensure that there exists a non-trivial solution $\mathbf{B}$ for \eqref{eqn:ZF_mat}, the dimension of $\mathcal{N}(\mathbf{H}_{\Nc_e,\Tc})$ should be greater than zero or $rank(\mathbf{H}_{\Nc_e,\Tc})\leq n_r$. Considering the worst-case scenario when $\mathbf{H}_{\Nc_e,\Tc}$ is a full rank matrix, the ZF strategy requires $n_e\leq n_r$. This condition is implied by the cardinality of the set of transmitters in \eqref{eqn:def_Tx_par}. This means that to combat eavesdroppers, one needs at least the same number of nodes as relays in this scheme.
Observing that the total power constraint \eqref{eqn:ach_ZF_cond_pow} is already obtained in Lemma~\ref{lemma:Ach_par_DF} completes the proof.
\end{IEEEproof}

\begin{figure}[tb]
  \centering
  \includegraphics[width=10cm]{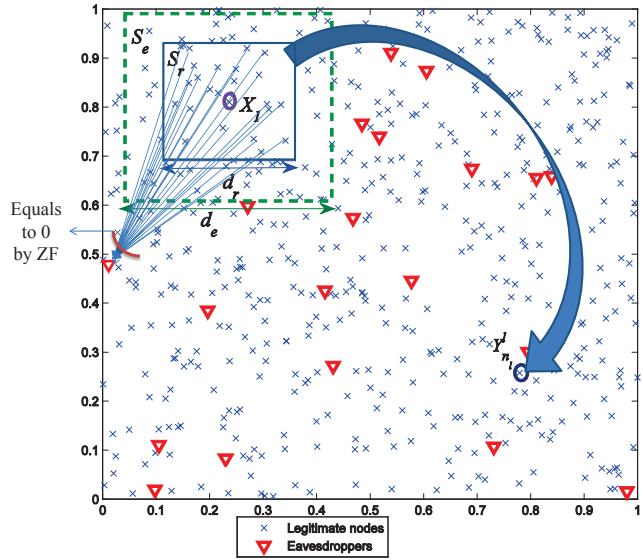}
  \caption{Parallel relaying illustrated for a typical network model, a square of unit area; with 500 legitimate nodes and 20 eavesdroppers, placed according to PPP. The relaying square $S_r$ of side $d_r$ is shown with solid line and the eavesdropper-free square $S_e$ of side $d_e$ is shown with dashed line. The source (with channel input $X_1$) is at the center of $S_r$ and $S_e$. We choose $\Tc$ as $n_r+1$ nodes in $S_r$. The destination node is shown with channel output $Y^l_{n_l}$. For brevity, ZF is only shown (solid lines originating $S_r$) in one eavesdropper.}
  \label{fig:par_points}
\end{figure}
\textbf{Step 2:} We start by choosing two random nodes in the network as our source-destination pair. Recall that $n_l,n_e\rightarrow\infty$. By applying Lemma~\ref{lemma:Poisson_che} (stated in Appendix~\ref{app:lemma:Poisson_che}), $n_l$ and $n_e$ can be made arbitrarily close to $\lambda_l$ and $\lambda_e$, respectively, with high probability (w.h.p). We define the \emph{relaying} square $S_r$ of side $d_r$, with the source at its center, as well as the \emph{eavesdropper-free} square $S_e$ of side $d_e$ (illustrated in Fig.~\ref{fig:par_points}) such that:
\begin{IEEEeqnarray}{c}
d_r=\sqrt{\frac{n_r}{n_l}} , d_e=\sqrt{\frac{n_r}{n_l}}(\log{n_e})^{\frac{\gamma}{2}} \mbox{ for some } \gamma>0.\label{eqn:par_squ}
\end{IEEEeqnarray}

The next lemma shows the feasibility of these squares.
\begin{lemma}\label{lemma:par_squ_nodes}
As long as $n_l\geq n_en_r(\log n_e)^{\gamma+\delta}$ for some $\gamma,\delta>0$, the probability of having at least $n_r$ legitimate nodes in $S_r$ tends to 1, and the probability of having the eavesdropper-free square $S_e$ can be made arbitrarily close to~1.
\end{lemma}
\begin{IEEEproof}
Using the fact that a Poisson process has Poisson increments, the number of nodes in $S_r$ is a two-dimensional Poisson RV with parameter $\lambda_ld_r^2$, at least equals to $n_r$ w.h.p (by applying \eqref{eqn:par_squ} and \eqref{eqn:Poisson_che-}) as long as $\lambda_ld_r^2\simeq n_r\rightarrow\infty$. This always holds because $n_r\geq n_e$. Similarly, the number of eavesdroppers in $S_e$ is a Poisson RV with parameter $\lambda_ed_e^2\simeq \frac{n_en_r}{n_l}(\log n_e)^\gamma$, which converges to 0 by applying the condition stated in this lemma. Hence, the probability of having no eavesdropper in $S_e$, i.e., $e^{-\lambda_ed_e^2}$, can be made arbitrarily close to~1.
\end{IEEEproof}
\textbf{Step 3:} Note that the number of relays, specified in the above lemma as $\frac{n_l}{n_e(\log n_e)^{\gamma+\delta}}$, should not be less than $n_e$. Now, we state the main result of this section and prove that the scaling of the nodes satisfies this constraint.
\begin{theorem}\label{thm:SL_par}
In $\Sc\Nc$ with fixed $\overline{P}_{tot}$ in \eqref{eqn:tot_power_cons}, as long as $n_e^{\frac{\alpha}{2}+1}(\log(n_e))^{\gamma+\delta(\frac{\alpha}{2}+1)}=o(n_l)$  holds for some positive $\gamma,\delta$, w.h.p. an infinite secure aggregate rate $\Rc_{s}$ is achievable.
\end{theorem}
\begin{IEEEproof}
First, we randomly choose the source of the message and call it node 1. According to Lemma~\ref{lemma:par_squ_nodes}, squares $S_r$ and $S_e$ with sides defined in \eqref{eqn:par_squ} exist w.h.p, with the source at their center. We randomly choose the destination and call it node $n_l$. If the destination is inside $S_r$, then the message is sent directly and no cooperation is needed. The model reduces to a wiretap channel with many eavesdroppers; the following rate, using Wyner wiretap coding at the source, is achievable:
\begin{IEEEeqnarray*}{rcl}
\Rc_s^{WT}&=&\min\limits_{j\in\Nc_e}\log(\frac{N^e}{N^l}\frac{N^l+|h_{1,n_l}^l|^2\tilde{P}_1}{N^e+|h_{1,j}^e|^2\tilde{P}_1})\yesnumber\label{eqn:Ach_WT}\\
&\stackrel{(a)}{\geq}&\log(\frac{N^e}{N^l}\frac{N^l+d_{r}^{-\alpha}\overline{P}_{tot}}{N^e+(\frac{d_{e}}{2})^{-\alpha}\overline{P}_{tot}})\log(\frac{d_e}{d_r})^{\alpha}\stackrel{(b)}{\rightarrow}\infty\quad \mbox{as}\quad n_l\rightarrow\infty
\end{IEEEeqnarray*}
where (a) is obtained by considering \eqref{eqn:chgain}, the definitions of $S_r$ and $S_e$ and by applying \eqref{eqn:tot_power_cons}, and (b) holds due to \eqref{eqn:par_squ}.
Otherwise, if the destination node is not in $S_r$, Lemma~\ref{lemma:par_squ_nodes} implies that w.h.p. we can construct the set of transmitters in \eqref{eqn:def_Tx_par}. Now, to make ZF possible we must show that $n_r=|\Tc|-1 \geq n_e$. By applying the constraint of Lemma~\ref{lemma:par_squ_nodes} with equality, we have
\begin{IEEEeqnarray*}{c}
n_r= \frac{n_l}{n_e(\log n_e)^{\gamma+\delta}}\stackrel{(a)}{=}n_e^{\frac{\alpha}{2}}(\log(n_e))^{\frac{\alpha}{2}\delta}\frac{n_l}{o(n_l)}\stackrel{(b)}{\geq} n_e
\end{IEEEeqnarray*}
(a) is due to the scaling condition stated in this theorem and (b) is obtained because $\alpha\geq 2$. Now, we can use the strategy of Theorem~\ref{thm:Ach_par_DF} to achieve \eqref{eqn:Ach_par_ZF}. To apply the total power constraint \eqref{eqn:ach_ZF_cond_pow}, in this case, we choose a fixed $\tilde{P}_1=\overline{P}_1$ and set $\tilde{P}_u=\frac{\overline{P}_{tot}-\overline{P}_1}{\|\mathbf{B}\|_2^2}$. First, we consider the first term in \eqref{eqn:Ach_par_ZF}, known as broadcast term (the secure rate from the source to $n_r$ relay nodes in $S_r$), and derive its asymptotic behavior as
\begin{IEEEeqnarray*}{rcl}
\log(\frac{N^e}{N^l}\frac{N^l+|h_{1,i^*}^l|^2\tilde{P}_1}{N^e+|h_{1,j}^e|^2\tilde{P}_1})&\stackrel{(a)}{\geq}&\log(\frac{N^e}{N^l}\frac{N^l+d_{r}^{-\alpha}\overline{P}_{1}}{N^e+(\frac{d_{e}}{2})^{-\alpha}\overline{P}_{1}})\\
&=&\log(\frac{d_e}{d_r})^{\alpha}\stackrel{(b)}{\rightarrow}\infty\quad \mbox{as}\quad n_l\rightarrow\infty\yesnumber\label{eqn:SL_par_1}
\end{IEEEeqnarray*}
where (a) is obtained by considering \eqref{eqn:chgain} and the defined squares, and (b) is due to \eqref{eqn:par_squ}. As expected, the rate to each node in $S_r$ is similar to the case where the destination is also in $S_r$; it can be made arbitrary large by decreasing the size of $S_r$ as needed. Note that this decrease needs larger $\lambda_l$ to have $n_r\geq n_e$ legitimate nodes in $S_r$ to employ them as relays. Before continuing to the second term in \eqref{eqn:Ach_par_ZF}, we take a closer look at the beamforming vector $\mathbf{B}\in \mathcal{N}(\mathbf{H}_{\Nc_e,\Tc})$. By applying Singular Value Decomposition (SVD), we have 
\begin{IEEEeqnarray*}{c}
\mathbf{H}_{\Nc_e,\Tc}=\mathbf{U}\mathbf{\Lambda}[\mathbf{\Upsilon}\mathbf{V}]^T;
\end{IEEEeqnarray*}
$\mathbf{\Upsilon}\in\mathbb{C}^{(n_r+1)\times n_e}$ contains the first $n_e$ right singular vectors corresponding to non-zero singular values, and $\mathbf{V}\in\mathbb{C}^{(n_r+1)\times(n_r-n_e+1)}$ contains the last $n_r-n_e+1$ singular vectors corresponding to zero singular values of $\mathbf{H}_{\Nc_e,\Tc}$. The later forms an orthonormal basis for the null space of $\mathbf{H}_{\Nc_e,\Tc}$. Hence, $\mathbf{B}$ can be expressed as their linear combination, i.e.,
\begin{IEEEeqnarray*}{c}
\mathbf{B}=\mathbf{V}\mathbf{\Phi}\label{eqn:null_mat_cond}
\end{IEEEeqnarray*}
where $\mathbf{\Phi}\in\mathbb{C}^{(n_r-n_e+1)}$ is an arbitrary vector selected by considering the power constraints in \eqref{eqn:ach_ZF_cond_pow}. Now, we consider the second term of \eqref{eqn:Ach_par_ZF}, known as the multi-access term. This corresponds to the cooperative secure rate from the source and the $n_r$ relays toward the destination.
\begin{IEEEeqnarray*}{rcl}
\max_{\mathbf{B}}\log(\frac{N^e}{N^l}\frac{N^l+|h_{1,{n_l}}^l|^2\tilde{P}_1+|\sum\limits_{k\in\Tc}h_{k,{n_l}}^l\beta_k|^2\tilde{P}_u}{N^e+|h_{1,j}^e|^2\tilde{P}_1})
&\stackrel{(a)}{\geq}&\max_{\mathbf{B}}\log(\frac{2^{\alpha/4}N^e}{N^l}\frac{|\mathbf{1}^\dag\mathbf{B}|^2\cdot\frac{\overline{P}_{tot}-\overline{P}_1}{\|\mathbf{B}\|_2^2}}{N^e+(\frac{d_{e}}{2})^{-\alpha}\overline{P}_{1}})\\
&=&\max_{\mathbf{\Phi}\dag\mathbf{\Phi}\leq \|\mathbf{B}\|_2^2} \log(\frac{2^{\alpha/4}N^e}{N^l}\frac{\mathbf{\Phi}^\dag\mathbf{V}^\dag\mathbf{1}\mathbf{1}^\dag\mathbf{V}\mathbf{\Phi}\cdot\frac{\overline{P}_{tot}-\overline{P}_1}{\|\mathbf{B}\|_2^2}}{N^e+(\frac{d_{e}}{2})^{-\alpha}\overline{P}_{1}})\\
&=&\log(\frac{2^{\alpha/4}N^e}{N^l}\frac{\lambda_{max}(\mathbf{V}^\dag\mathbf{1}\mathbf{1}^\dag\mathbf{V})\cdot(\overline{P}_{tot}-\overline{P}_1)}{N^e+(\frac{d_{e}}{2})^{-\alpha}\overline{P}_{1}})\\
&=&\log(\frac{2^{\alpha/4}N^e}{N^l}\frac{\|\mathbf{1}^\dag\mathbf{V}\|^2_2\cdot(\overline{P}_{tot}-\overline{P}_1)}{N^e+(\frac{d_{e}}{2})^{-\alpha}\overline{P}_{1}})\\
&\stackrel{(b)}{=}&\log(\frac{2^{\alpha/4}N^e}{N^l}\frac{(n_r+1)\frac{1+\cos2\theta}{2}\cdot(\overline{P}_{tot}-\overline{P}_1)}{N^e+(\frac{d_{e}}{2})^{-\alpha}\overline{P}_{1}})\\
&\stackrel[n_l\rightarrow\infty]{(c)}{=}&\log(\kappa\frac{(n_r+1)}{d_{e}^{-\alpha}})\stackrel{(d)}{\rightarrow}\infty\quad \mbox{as}\quad n_l\rightarrow\infty\yesnumber\label{eqn:SL_par_22}
\end{IEEEeqnarray*}
(a) holds since $d_{k,{n_l}}^l\leq\sqrt{2}$ and $\mathbf{1}\in\mathbb{C}^{(n_r+1)}$ is the all one vector. In (b), $\theta$ is an RV denotes the angle between $\mathbf{1}$ and $\mathcal{N}(\mathbf{H}_{\Nc_e,\Tc})$ and has a continuous distribution on $[0,2\pi]$ due to the randomness of $\mathbf{H}_{\Nc_e,\Tc}$. In (c), $\kappa$ is a constant. (d) is obtained by substituting \eqref{eqn:par_squ} and $n_r= \frac{n_l}{n_e(\log n_e)^{\gamma+\delta}}$ and by applying the scaling $n_e^{\frac{\alpha}{2}+1}(\log(n_e))^{\gamma+\delta(\frac{\alpha}{2}+1)}=o(n_l)$ for some positive $\gamma,\delta$. This completes the proof.
\end{IEEEproof}

\section{Serial Relaying}\label{sec:Ser}
In this section, we improve the scaling of the number of eavesdroppers we can defend against at the expense of a more complicated strategy, serial (multi-stage) relaying. The network is divided into clusters, with the nodes in each cluster acting as a group of relays and, at the same time, collectively applying ZF (essentially acting as a distributed multi-antenna). These clusters perform \emph{ordered} DF: the nodes in each cluster decode the transmitted signals of all previous clusters. We use the three-step approach outlined in Section~\ref{sec:Par} to obtain our result here. We show that unbounded secure aggregate rate for a fixed total power can be achieved as long as $n_e^2(\log(n_e))^{\gamma}=o(n_l)$ holds for some positive $\gamma$.

\textbf{Step 1:} Achievability is given in the following theorem.
\begin{theorem}\label{thm:Ach_ser_DF}
For $\Sc\Nc$, the following secure aggregate rate is achievable:
\begin{IEEEeqnarray}{rcl}
\Rc_s^{DF,ZF,ser}&=&\min\limits_{i\in[1:n_l-1]}\max_{\mathbf{B}_i,\tilde{P}_i}\min\limits_{j\in\Nc_e}\log(\frac{N^e}{N^l}\frac{N^l+\sum\limits_{q=1}^i|\sum\limits_{k=1}^qh_{k,{i+1}}^l\beta'_{kq}|^2\tilde{P}_q}{N^e+|h_{1,j}^e|^2\tilde{P}_1})\label{eqn:Ach_ser_ZF}
\end{IEEEeqnarray}
in which
\begin{IEEEeqnarray}{l}
\beta'_{kq}=\mathbf{B}_q(k)\quad\mbox{and}\quad \beta'_{kq}=1 \quad\mbox{if}\quad k=q\label{eqn:ach_coef_ser}\\
\mathbf{B}_q\in {\mathcal{N}(\mathbf{H}_{\Nc_e,\Tc^q})} \quad\mbox{for}\quad q\bmod n_e=1\label{eqn:ach_ZF_cond_coef_ser}\\
\tilde{P}_q = \left\{ \begin{array}{ll}
\overline{P}_q & \textrm{if }q\bmod n_e=1\\
0 & \textrm{if }q\bmod n_e\neq1
\end{array} \right.\label{eqn:ach_ZF_cond_cluspow_ser}\\
\sum\limits_{q=1}^{n_l-1}\|\mathbf{B}_q\|_2^2\tilde{P}_q\leq \overline{P}_{tot}\label{eqn:ach_ZF_cond_pow_ser}
\end{IEEEeqnarray}
where $\mathbf{H}_{\Nc_e,\Tc^q}\in\mathbb{C}^{n_e\times q}$ is the cluster-eavesdroppers channel matrix which its $(j,i)$th element is $h_{i,j}^e$ for $i\in[1:q],j\in\Nc_e$.
\end{theorem}
\begin{IEEEproof}
We use a $(n_l-1)$-stage block Markov coding by making the nodes relaying the message with ordered priorities. Considering the ordered set for the legitimate nodes, i.e., $\Nc_l=\{1,\ldots,n_l\}$, each node $i$ decodes the transmitted signal of all previous nodes (1 to $i-1$) in this order and sends its signal to the subsequent nodes. In order to pipeline communication, $(n_l-1)$th order block Markov correlated codes are proposed. Therefore, in each block, the received signals at the legitimate nodes are coherent \cite{KraGasGup05}. To apply ZF at the eavesdroppers, we show it is necessary to have clusters of enough relays, where the nodes in each cluster have the same priority compared to the source. Wyner wiretap coding is also utilized at the source. First, we use the multi-stage block Markov coding (serial DF relaying) and Wyner wiretap coding to obtain $\Rc_s^{DM,ser}$ in Lemma~\ref{lemma:Ach_ser_DF_DM} (proof provided in Appendix~\ref{app:proof_lemma:Ach_ser_DF_DM}) and extend it to $\Rc_s^{DF,ser}$ for the Gaussian channel in Lemma~\ref{lemma:Ach_ser_DF} (proof in Appendix~\ref{app:proof_lemma:Ach_ser_DF}). Then, by applying ZF on $\Rc_s^{DF,ser}$ we derive $\Rc_s^{DF,ZF,ser}$.
\begin{lemma}\label{lemma:Ach_ser_DF_DM}
Consider the channel model of Lemma~\ref{lemma:Ach_par_DF_DM} and let $\pi(\cdot)$ be a permutation on $\Nc_l=\{1,\ldots,n_l\}$, where $\pi(1)=1$, $\pi(n_l)=n_l$ and $\pi(m:n)=\{\pi(m),\pi(m+1),\ldots,\pi(n)\}$. The secrecy capacity is lower-bounded by:
\begin{IEEEeqnarray}{l}\label{eqn:Ach_ser_DF_DM}
\Rc_s^{DM,ser}=\sup\min\limits_{j\in\Nc_e}\max\limits_{\pi(\cdot)}\min\limits_{i\in[1:n_l-1]}I(U_{\pi(1:i)};Y_{\pi(i+1)}^l|U_{\pi(i+1:n_l-1)})-I(U_{\pi(1:n_l-1)};Y_{j}^e)\nonumber
\end{IEEEeqnarray}
where the supremum is taken over all joint p.m.fs of the form
\begin{IEEEeqnarray}{c}\label{eqn:Ach_ser_DF_DM_pmf}
p(u_1,\ldots,u_{n_l-1})\prod\limits_{k=1}^{n_l-1} p(x_k|u_k).
\end{IEEEeqnarray}
\end{lemma}
Similar to the parallel relaying case, we choose an appropriate suboptimal input distribution in the following lemma.
\begin{lemma}\label{lemma:Ach_ser_DF}
For $\Sc\Nc$, the following is an achievable secure aggregate rate:
\begin{IEEEeqnarray*}{l}
\Rc_s^{DF,ser}=\min\limits_{i\in[1:n_l-1]}\max_{\mathbf{B}_i,\tilde{P}_i}\min\limits_{j\in\Nc_e}\log(1+\frac{\sum\limits_{q=1}^i|\sum\limits_{k=1}^qh_{k,{i+1}}^l\beta'_{kq}|^2\tilde{P}_q}{N^l})-\log(1+\frac{\sum\limits_{q=1}^{n_l-1}|\sum\limits_{k=1}^qh_{k,{j}}^e\beta'_{kq}|^2\tilde{P}_q}{N^e})\yesnumber\label{eqn:Ach_ser_DF}
\end{IEEEeqnarray*}
where \eqref{eqn:ach_coef_ser} and \eqref{eqn:ach_ZF_cond_pow_ser} hold.
\end{lemma}

For the serial relaying scheme, as the achievable rate is not limited by the decoding constraint at the farthest relay, all nodes in the network (except the source and destination) can be used as the relay nodes. Therefore, the transmission set can be $\Tc=\{1,\ldots,n_l-1\}$ where the relays are assumed to be in a certain order, e.g., based on their distances to the source node. Similar to Section~\ref{sec:Par}, we apply ZF at all eavesdroppers to determine the beamforming coefficient vectors $\mathbf{B}_q$ by setting $\sum\limits_{q=2}^{n_l-1}|\sum\limits_{k=1}^qh_{k,{j}}^e\beta'_{kq}|^2\tilde{P}_q=0,\forall j\in\Nc_e$. This results in $\tilde{P}_q=0$ or 
\begin{IEEEeqnarray*}{c}
E(q,j)=\sum\limits_{k=1}^qh_{k,{j}}^e\beta'_{kq}=0, \label{eqn:Ach_par_DF_ZF_cons}
\end{IEEEeqnarray*}
for $\forall q\in[2:n_l-1]$. Now consider \eqref{eqn:map2_Ach_ser_DF} to obtain $X_k=\tilde{U}_k+\beta_{k}X_{k+1}$ where $\beta'_{kq}=\prod\limits_{m=k}^{q-1}\beta_m$. Therefore, $E(q_0,j)$ and $E(q_0+1,j)$ only differ in one variable, i.e., $\beta_{q_0+1}$. However, we need $E(q,j)=0,\forall j\in\Nc_e$ if $\tilde{P}_q>0$, which is clearly not possible. Therefore, we apply ZF by allocating power as per \eqref{eqn:ach_ZF_cond_cluspow_ser} and $E(q,j)=0, \textrm{if }q\bmod n_e=1, \forall j\in\Nc_e$. Thus, we obtain $\mathbf{H}_{\Nc_e,\Tc^q}\mathbf{B}_q=\mathbf{0}$ shown in \eqref{eqn:ach_ZF_cond_coef_ser} ($\mathbf{H}_{\Nc_e,\Tc^q}$ is defined in Theorem~\ref{thm:Ach_ser_DF}). By applying \eqref{eqn:ach_ZF_cond_coef_ser} on \eqref{eqn:Ach_ser_DF}, we achieve \eqref{eqn:Ach_ser_ZF}. This means that to overcome $n_e$ eavesdroppers using the proposed strategy, one node in every $n_e$ legitimate nodes should transmit fresh information and the $n_e$ nodes who transmit the same information in each block should apply beamforming to ZF at all eavesdroppers.
\end{IEEEproof}

\begin{figure}[tb]
  \centering
  \includegraphics[width=10cm]{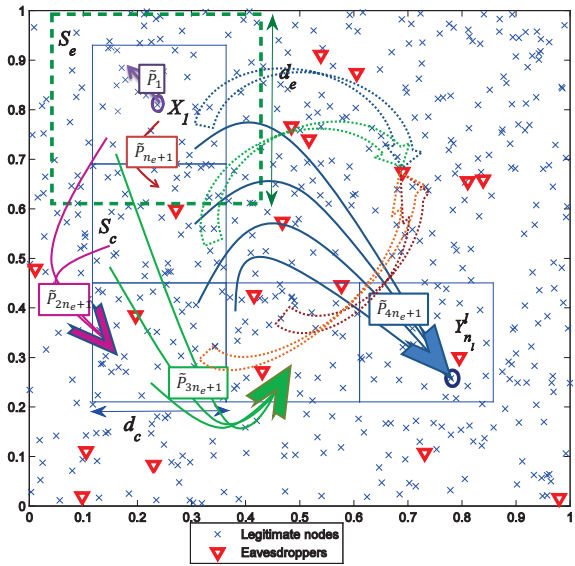}
  \caption{Clusters (squares $S_c$, thin solid line) of side $d_c$ used for serial relaying. The nodes in each cluster, $c$, coherently with nodes in all previous clusters and $i-c$ subsequent clusters send $\tilde{P}_{in_e+1}$ to the nodes in cluster $i+1$ for $i\geq c$. Each dotted arrow shows the received signals at one eavesdropper from all nodes in each cluster; these are equal to zero thanks to ZF.}
  \label{fig:ser_points}
\end{figure}
\textbf{Step 2:}
Consider Fig.~\ref{fig:ser_points} and assume $c_{max}$ clusters (squares) $S_c$ of same side $d_c$; the source is located in the first and the destination in the last cluster; any two successive clusters share one side. Hence, we have $\frac{1}{d_c}\leq c_{max}\leq \frac{2}{d_c}$. We show $c_{max}$ does not affect the asymptotic behavior of $\Rc_s$. Assuming the strategy of Step~1, all the nodes in each cluster $S_c$ transmit the same fresh information. Now, consider an eavesdropper-free square $S_e$ of side $d_e$ around the source. We define
\begin{IEEEeqnarray}{c}
d_c=\sqrt{\frac{n_c}{n_l}} , d_e=\sqrt{\frac{n_c}{n_l}}(\log{n_e})^{\frac{\gamma}{2}}\mbox{ for some } \gamma>0\label{eqn:ser_squ}
\end{IEEEeqnarray}
where $n_c$ is determined in  Lemma~\ref{lemma:ser_squ_nodes} which shows the feasibility of these squares and can be proved similar to Lemma~\ref{lemma:par_squ_nodes}.
\begin{lemma}\label{lemma:ser_squ_nodes}
As long as $n_en_c(\log n_e)^{\gamma}=o(n_l)$ for some $\gamma>0$, the probability of having at least $n_c\rightarrow\infty$ legitimate nodes in $S_r$ tends to 1, and the probability of having no eavesdropper in square $S_e$ can be made arbitrarily close to~1.
\end{lemma}

We remark that to apply ZF at all eavesdroppers, the number of nodes in each cluster, i.e., $n_c$, should not be less than $n_e$.

\textbf{Step 3:} Now, we state the main result of this section.
\begin{theorem}\label{thm:SL_ser}
In $\Sc\Nc$ with fixed total power $\overline{P}_{tot}$ in \eqref{eqn:tot_power_cons}, as long as $n_e^2(\log n_e)^{\gamma}=o(n_l)$ holds for some positive $\gamma$, w.h.p. an infinite secure aggregate rate $\Rc_{s}$ is achievable.
\end{theorem}
\begin{IEEEproof}
Choosing the source and the destination is same as proof of Theorem~\ref{thm:SL_par}. If the destination is inside the square $S_c$, the message is sent directly to it using Wyner wiretap coding at the source. Similar to \eqref{eqn:Ach_WT}, since $\frac{d_e}{d_c}\rightarrow\infty$ as $n_l\rightarrow\infty$, in this case an unbounded rate is achievable. Otherwise (the destination is outside $S_c$), we choose $n_c=n_e+1$ and consider $c_{max}$ clusters as described in the previous step. Substituting $n_c=n_e+1$ into the scaling of Lemma~\ref{lemma:ser_squ_nodes} results in scaling of this theorem. As $n_c\geq n_e$, w.h.p ZF can be applied and the rate of Theorem~\ref{thm:Ach_ser_DF} is achievable. If we consider equal power allocation for the fresh information in the total power constraint \eqref{eqn:ach_ZF_cond_pow_ser}, we obtain 
\begin{IEEEeqnarray*}{l}
\tilde{P}_q=\frac{\overline{P}_{tot}}{\sum\limits_{c=0}^{c_{max}}\|\mathbf{B}_{cn_e+1}\|_2^2}=\overline{P}_q=\overline{P},
\end{IEEEeqnarray*}
if $q\bmod n_e=1$ and $q\leq c_{max}n_e+1$. Otherwise ($q\bmod n_e\neq1$), $\tilde{P}_q=0$. Note that we consider an ordered set of legitimate nodes based on the cluster numbers, which can be done w.h.p according to Lemma~\ref{lemma:ser_squ_nodes}. Now, we show that \eqref{eqn:Ach_ser_ZF} can be unbounded w.h.p for all $i\in[1:n_l-1],j\in\Nc_e$ and $\mathbf{B}_q$s that satisfy \eqref{eqn:ach_coef_ser} and \eqref{eqn:ach_ZF_cond_coef_ser}. First, we consider $i\leq n_e+1$ that comprises the nodes in the first cluster:
\begin{IEEEeqnarray*}{rcl}
\Rc_s^{DF,ZF,ser}&\stackrel{(a)}{=}&\log(\frac{N^e}{N^l}\frac{N^l+|h_{k,{i+1}}^l|^2\overline{P}_1}{N^e+|h_{1,j}^e|^2\overline{P}_1})\yesnumber\label{eqn:SL_ser_1}\\
&\stackrel{(b)}{\geq}&\log(\frac{N^e}{N^l}\frac{N^l+d_{c}^{-\alpha}\overline{P}_{1}}{N^e+d_e^{-\alpha}\overline{P}_{1}}){\rightarrow}\infty\quad \mbox{as}\quad n_l\rightarrow\infty
\end{IEEEeqnarray*}
where (a) is due to \eqref{eqn:ach_ZF_cond_cluspow_ser} and (b) is obtained by considering \eqref{eqn:chgain} and the defined squares in \eqref{eqn:ser_squ}. This rate is similar to the one we have in \eqref{eqn:SL_par_1}. In fact, one expects that this rate can be made arbitrary large if we choose $S_c$ small enough (by increasing the density of nodes). In parallel relaying, the problem with the second term in \eqref{eqn:Ach_par_ZF} is the fixed non-decreasing distance between the nodes in $S_r$ and the destination. We here overcome this problem by considering clusters such that the maximum distance between the nodes in two adjacent clusters is $\sqrt{5}d_c$. Therefore, for the nodes in cluster $c$, i.e., $cn_e+1\leq i\leq (c+1)n_e$, we set $q=cn_e+1$:
\begin{IEEEeqnarray*}{rcl}
\Rc_s^{DF,ZF,ser}&\geq&\log(\frac{N^e}{N^l}\frac{N^l+|\sum\limits_{k=1}^qh_{k,{i+1}}^l\beta'_{kq}|^2\tilde{P}_{q}}{N^e+|h_{1,j}^e|^2\tilde{P}_1})\\
&\stackrel{(a)}{=}&\log(\frac{N^e}{N^l}\frac{N^l+|\mathbf{h}_q^T\mathbf{B}_q|^2\overline{P}}{N^e+|h_{1,j}^e|^2\overline{P}})\\
&\stackrel[n_l\rightarrow\infty]{(b)}{\geq}&\log(\frac{d_e}{d_c})^{\alpha}\stackrel{(c)}{\rightarrow}\infty\quad \mbox{as}\quad n_l\rightarrow\infty\yesnumber\label{eqn:SL_ser_2}
\end{IEEEeqnarray*}
(a) is obtained by defining $\mathbf{h}_q=[h_{1,{i+1}}^l,\ldots,h_{q,{i+1}}^l]^T$. (b) follows from the steps similar to \eqref{eqn:SL_par_22} and from noting that $\|\mathbf{B}_q\|_2^2\geq \mathbf{B}_q(q)=\beta'_{qq}=1$, $\|\mathbf{h}_q\|_2^2\geq |h_{q,{i+1}}^l|^2\geq  d_c^{-\alpha}$ and the randomness of $\mathbf{H}_{\Nc_e,\Tc^q}$. (c) is due to \eqref{eqn:ser_squ}. This completes the proof.
\end{IEEEproof}

\section{Colluding Eavesdroppers}\label{sec:colluding}
In this section, we state the maximum number of perfect-colluding eavesdroppers that can be tolerated in a zero-cost secure communication using a relaying based scheme. In fact, we show that if $n_e^{(2+\frac{2}{\alpha})}(\log n_e)^{\gamma}=o(n_l)$ holds for some positive $\gamma$, we achieve an unbounded secure aggregate rate for a fixed total power.
For the proof, we adapt the framework of Section~\ref{sec:Ser} to the colluding case, where in each step, the collusion should be taken into consideration.

1) \emph{A lower bound to the secrecy capacity:} We propose an achievability scheme in Theorem~\ref{thm:Ach_ser_DF-c} for a multiple relay channel in presence of perfect \emph{colluding} eavesdroppers.

2) \emph{Fitting the achievability scheme of Step~1 to $\Sc\Nc$-PCE:} By choosing appropriate values for the parameters of the first step, the constraints on the number of legitimate nodes and eavesdroppers are derived in Lemma~\ref{lemma:ser_squ_nodes-c}, under which the achievability results of Theorem~\ref{thm:Ach_ser_DF-c} can be applied to $\Sc\Nc$-PCE.

3) \emph{Infinite secure aggregate rate:} We show that the achievable secure aggregate rate of the first step is unbounded after applying the fixed total power constraint (in Theorem~\ref{thm:SL_ser-c}). Hence, the maximum number of the tolerable perfect colluding eavesdroppers is obtained.

\textbf{Step 1: }The following theorem presents an achievable secure rate for a multiple relay channel in the presence of \emph{colluding} eavesdroppers. We use serial (multi-stage) active cooperation (relaying), randomized encoding and beamforming through ZF. To make the ZF possible, we divide the network into clusters, where the nodes in each cluster act as a group of relays and, at the same time, collectively apply ZF (essentially as a distributed multi-antenna) on the colluding eavesdroppers. Applying this strategy results in some conditions on the clustering (such as the number of the nodes in each cluster). The proof is provided in Appendix~\ref{app:proof_thm:Ach_ser_DF-c}.
\begin{theorem}\label{thm:Ach_ser_DF-c}
For $\Sc\Nc$-PCE, the following secure aggregate rate is achievable:
\begin{IEEEeqnarray}{rcl}
\!\!\!\!\!\!\!\!\!\Rc_s^{ZF}&=&\min\limits_{i\in[1:n_l-1]}\max_{\mathbf{B}_i,\tilde{P}_i}\log(\frac{N^e}{N^l}\frac{N^l+\sum\limits_{q=1}^i|\sum\limits_{k=1}^qh_{k,{i+1}}^l\beta'_{kq}|^2\tilde{P}_q}{N^e+\sum\limits_{j\in\Nc_e}|h_{1,j}^e|^2\tilde{P}_1})\label{eqn:Ach_ser_ZF-c}
\end{IEEEeqnarray}
in which
\begin{IEEEeqnarray}{l}
\beta'_{kq}=\mathbf{B}_q(k)\quad\mbox{and}\quad \beta'_{kq}=1 \quad\mbox{if}\quad k=q\label{eqn:ach_coef_ser-c}\\
\mathbf{B}_q\in {\mathcal{N}(\mathbf{H}_{\Nc_e,\Tc^q})} \quad\mbox{for}\quad q\bmod n_e=1\label{eqn:ach_ZF_cond_coef_ser-c}\\
\tilde{P}_q = \left\{ \begin{array}{ll}
\overline{P}_q & \textrm{if }q\bmod n_e=1\\
0 & \textrm{if }q\bmod n_e\neq1
\end{array} \right.\label{eqn:ach_ZF_cond_cluspow_ser-c}\\
\sum\limits_{q=1}^{n_l-1}\|\mathbf{B}_q\|_2^2\tilde{P}_q\leq \overline{P}_{tot}\label{eqn:ach_ZF_cond_pow_ser-c}
\end{IEEEeqnarray}
where $\mathbf{H}_{\Nc_e,\Tc^q}\in\mathbb{C}^{n_e\times q}$ is the cluster-eavesdroppers channel matrix; its $(j,i)$th element is $h_{i,j}^e$, for $i\in[1:q],j\in\Nc_e$.
\end{theorem}

\begin{figure}[tb]
  \centering
  \includegraphics[width=9cm]{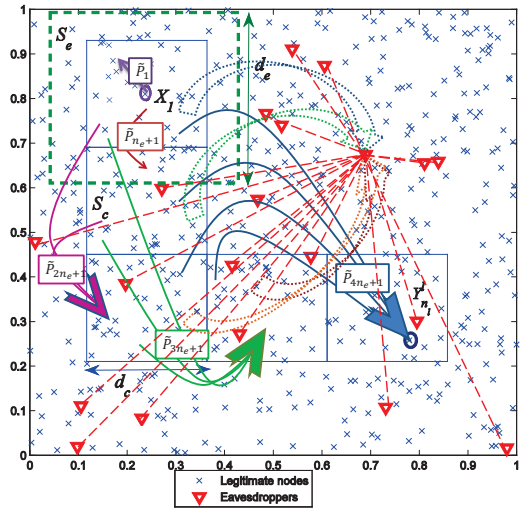}
  \caption{The squares are same as the ones in Fig.~\ref{fig:ser_points}. Here, the dashed lines (only shown for one eavesdropper) show the free access of the eavesdroppers to all observations.}
  \label{fig:ser_points-c}
\end{figure}

\textbf{Step 2:} Now, we specify the details of our strategy and derive the constraints on the number of legitimate nodes and eavesdroppers in Lemma~\ref{lemma:ser_squ_nodes-c} (to apply the scheme of Theorem~\ref{thm:Ach_ser_DF-c} to $\Sc\Nc$-PCE). First, we choose randomly the source-destination pair in $\Nc_l$. Since $n_l,n_e\rightarrow\infty$, we can apply Lemma~\ref{lemma:Poisson_che} (stated in Appendix~\ref{app:lemma:Poisson_che}) to make $n_l$ and $n_e$ arbitrarily close to $\lambda_l$ and $\lambda_e$, respectively, with high probability (w.h.p). We design $c_{max}$ clusters (squares), $S_c$, of same side $d_c$ (as shown in Fig.~\ref{fig:ser_points-c}); we consider an ordered set of nodes in clusters, with the source in the first cluster and the destination in the last one; any two successive clusters share one side. This results in $\frac{1}{d_c}\leq c_{max}\leq \frac{2}{d_c}$. In fact, the following results show that the asymptotic behavior of $\Rc_s$ is independent of $c_{max}$. Adapting the strategy of Step~1, each cluster ($S_c$) consists of the nodes transmitting the same part of the fresh information: in each cluster, only one node transmits fresh information. We only need one eavesdropper-free square, $S_e$, of side $d_e$ around the source (the remaining communications are secured through beamforming). We define:
\begin{IEEEeqnarray}{c}
d_c=\sqrt{\frac{n_c}{n_l}} , d_e=n_e^{\frac{1}{\alpha}}\sqrt{\frac{n_c}{n_l}}(\log{n_e})^{\frac{\gamma}{2}}\mbox{ for some } \gamma>0\label{eqn:ser_squ-c}
\end{IEEEeqnarray}
where $n_c$ is given in the following lemma that shows the feasibility of designing these squares.
\begin{lemma}\label{lemma:ser_squ_nodes-c}
If $n_cn_e^{(1+\frac{2}{\alpha})}(\log n_e)^{\gamma}=o(n_l)$ holds for some $\gamma>0$, the probability of having at least $n_c\rightarrow\infty$ legitimate nodes in $S_c$ goes to 1, and the probability of having no eavesdropper in square $S_e$ approaches~1.
\end{lemma}
\begin{IEEEproof}
Using the fact that a Poisson process has Poisson increments, the \emph{number} of nodes in $S_c$ is a two-dimensional Poisson RV with parameter $\lambda_ld_c^2$.
As long as $\lambda_ld_c^2\simeq n_c\rightarrow\infty$ holds, we can apply Lemma~\ref{lemma:Poisson_che} (stated in Appendix~\ref{app:lemma:Poisson_che}) on \eqref{eqn:ser_squ-c} to show that this number is greater than $n_c$ w.h.p. Recall that  to apply ZF at all eavesdroppers, we need at least $n_e$ nodes in each cluster, i.e., $n_c\geq n_e$. Thus, the above condition already holds ($n_c\geq n_e\rightarrow\infty$).
The number of eavesdroppers in $S_e$ is also a Poisson RV. Considering \eqref{eqn:ser_squ-c} and the condition stated in this lemma, we derive the parameter of this RV as: $\lambda_ed_e^2\simeq \frac{n_cn_e^{(1+\frac{2}{\alpha})}}{n_l}(\log n_e)^\gamma\rightarrow 0$. Now, the probability of having no eavesdropper in $S_e$ equals to $e^{-\lambda_ed_e^2}\rightarrow 1$. This completes the proof.
\end{IEEEproof}

\textbf{Step 3:} Now, we state our main result of this section.
\begin{theorem}\label{thm:SL_ser-c}
Considering the fixed total power ($\overline{P}_{tot}$) constraint in \eqref{eqn:tot_power_cons} for $\Sc\Nc$-PCE, an infinite secure aggregate rate $\Rc_{s}$ is achievable (w.h.p.), as long as $n_e^{(2+\frac{2}{\alpha})}(\log n_e)^{\gamma}=o(n_l)$ holds for some positive $\gamma$.
\end{theorem}
\begin{IEEEproof}
Randomly choose the source-destination pair; let the source be node 1 and the destination node $n_l$; design the squares $S_c$ and $S_e$ as per \eqref{eqn:ser_squ-c} (around the source), which exist w.h.p due to Lemma~\ref{lemma:ser_squ_nodes-c}. Moreover, design the clusters with an ordered set of legitimate nodes (based on the cluster numbers), which is feasible w.h.p according to Lemma~\ref{lemma:ser_squ_nodes-c}. Consider the following cases:

\emph{Case~1: the destination is inside the first cluster ($S_c$).} The source directly sends its message to the destination without any cooperation. In fact, all other nodes are silent. Therefore, the network reduces to a wiretap channel with many perfect colluding eavesdroppers. We use Wyner wiretap coding at the source to achieve the following unbounded rate:
\begin{IEEEeqnarray*}{rcl}
\Rc_s^{WT}&=&\log(\frac{N^e}{N^l}\frac{N^l+|h_{1,n_l}^l|^2\tilde{P}_1}{N^e+\sum\limits_{j\in\Nc_e}|h_{1,j}^e|^2\tilde{P}_1})\yesnumber\label{eqn:Ach_WT-c}\\
&\stackrel{(a)}{\geq}&\log(\frac{N^e}{N^l}\frac{N^l+d_{c}^{-\alpha}\overline{P}_{tot}}{N^e+n_e(\frac{d_{e}}{2})^{-\alpha}\overline{P}_{tot}})\stackrel{(b)}{\rightarrow}\infty\quad \mbox{as}\quad n_l\rightarrow\infty
\end{IEEEeqnarray*}
where (a) follows from \eqref{eqn:chgain} and \eqref{eqn:tot_power_cons} (by considering the concepts of $S_c$ and $S_e$ squares); (b) follows from \eqref{eqn:ser_squ-c}.

\emph{Case~2: the destination is outside the first cluster ($S_c$).}  Now, design the previously described $c_{max}$ clusters each with $n_c=n_e+1$ nodes. By substituting $n_c=n_e+1$ into the scaling of Lemma~\ref{lemma:ser_squ_nodes-c}, one can obtain $n_e^{(2+\frac{2}{\alpha})}(\log n_e)^{\gamma}=o(n_l)$, i.e., the scaling of this theorem. As $n_c\geq n_e$, w.h.p, ZF can be applied and we achieve the rate of Theorem~\ref{thm:Ach_ser_DF-c}. Now, we allocate the power equally to the fresh information based on the total power constraint \eqref{eqn:ach_ZF_cond_pow_ser-c}. Thus, we have $\tilde{P}_q=\frac{\overline{P}_{tot}}{\sum\limits_{c=0}^{c_{max}}\|\mathbf{B}_{cn_e+1}\|_2^2}=\overline{P}_q=\overline{P}$,
if $q\bmod n_e=1$ and $q\leq c_{max}n_e+1$. Otherwise, if $q\bmod n_e\neq1$), we set $\tilde{P}_q=0$. Thus, we can substitute these allocations into \eqref{eqn:Ach_ser_ZF-c} and investigate its asymptotic behavior for all $i\in[1:n_l-1]$ and $\mathbf{B}_q$s that satisfy \eqref{eqn:ach_coef_ser-c} and \eqref{eqn:ach_ZF_cond_coef_ser-c}. First, we consider the nodes in the first cluster by letting $i\leq n_e+1$:
\begin{IEEEeqnarray*}{rcl}
\Rc_s^{ZF}&\stackrel{(a)}{=}&\log(\frac{N^e}{N^l}\frac{N^l+|h_{k,{i+1}}^l|^2\overline{P}_1}{N^e+\sum\limits_{j\in\Nc_e}|h_{1,j}^e|^2\overline{P}_1})\yesnumber\label{eqn:SL_ser_1_c}\\
&\stackrel{(b)}{\geq}&\log(\frac{N^e}{N^l}\frac{N^l+d_{c}^{-\alpha}\overline{P}_{1}}{N^e+n_ed_e^{-\alpha}\overline{P}_{1}}){\rightarrow}\infty\quad \mbox{as}\quad n_l\rightarrow\infty
\end{IEEEeqnarray*}
(a) follows from the power allocation as in \eqref{eqn:ach_ZF_cond_cluspow_ser-c}. (b) follows from the network model in \eqref{eqn:chgain} and the clustering (squares) concept with the sizes as per \eqref{eqn:ser_squ-c}. The intuition is to make the cluster $S_c$ small enough to increase the rate achievable toward the nodes in the first cluster (similar to Case~1). However, by this reduction in the cluster size, one needs larger $\lambda_l$ to have enough nodes in each cluster to make ZF possible at all eavesdroppers (i.e., $n_c\geq n_e$). This trade-off specifies the scaling.

Now, before continuing to the rate of the other clusters, let us take a closer look at the beamforming vector of each cluster ($\mathbf{B}_q\in {\mathcal{N}(\mathbf{H}_{\Nc_e,\Tc^q})}$ for $q\bmod n_e=1$). By applying Singular Value Decomposition (SVD), we have $\mathbf{H}_{\Nc_e,\Tc^q}=\mathbf{U}_q\mathbf{\Lambda}_q[\mathbf{\Upsilon}_q\mathbf{V}_q]^T$;
$\mathbf{\Upsilon}_q\in\mathbb{C}^{q\times n_e}$ contains the first $n_e$ right singular vectors corresponding to non-zero singular values, and $\mathbf{V}_q\in\mathbb{C}^{q\times(q-n_e)}$ contains the last $q-n_e$ singular vectors corresponding to zero singular values of $\mathbf{H}_{\Nc_e,\Tc^q}$. The later forms an orthonormal basis for the null space of $\mathbf{H}_{\Nc_e,\Tc^q}$. Hence, $\mathbf{B}_q$ can be expressed as their linear combination, i.e., $\mathbf{B}_q=\mathbf{V}_q\mathbf{\Phi}_q$,
where $\mathbf{\Phi}_q\in\mathbb{C}^{(q-n_e)}$ is an arbitrary vector selected by considering the power constraints in \eqref{eqn:ach_ZF_cond_pow_ser-c}.

Now, consider the nodes in cluster $c$, i.e., $cn_e+1\leq i\leq (c+1)n_e$, and set $q=cn_e+1$. We remark that to overcome the fixed, non-decreasing distance between the nodes in the first cluster and the destination, the clusters are designed such that the maximum distance between the nodes in two adjacent clusters is $\sqrt{5}d_c$.
\begin{IEEEeqnarray*}{rcl}
\Rc_s^{ZF}&=&\max_{\mathbf{B}_i}\log(\frac{N^e}{N^l}\frac{N^l+|\sum\limits_{k=1}^qh_{k,{i+1}}^l\beta'_{kq}|^2\tilde{P}_{q}}{N^e+\sum\limits_{j\in\Nc_e}|h_{1,j}^e|^2\tilde{P}_1})\\
&\stackrel{(a)}{=}&\max_{\mathbf{B}_q}\log(\frac{N^e}{N^l}\frac{N^l+|\mathbf{h}_q^\dag\mathbf{B}_q|^2\overline{P}}{N^e+\sum\limits_{j\in\Nc_e}|h_{1,j}^e|^2\overline{P}})\\
&{=}&\max_{\mathbf{\Phi_q}\dag\mathbf{\Phi_q}\leq \|\mathbf{B}_q\|_2^2} \log(\frac{N^e}{N^l}\frac{N^l+\mathbf{\Phi}_q^\dag\mathbf{V}_q^\dag\mathbf{h}_q\mathbf{h}_q^\dag\mathbf{V}_q\mathbf{\Phi}_q\overline{P}}{N^e+\sum\limits_{j\in\Nc_e}|h_{1,j}^e|^2\overline{P}})\\
&{=}& \log(\frac{N^e}{N^l}\frac{N^l+\|\mathbf{B}_q\|_2^2\lambda_{max}(\mathbf{V}_q^\dag\mathbf{h}_q\mathbf{h}_q^\dag\mathbf{V}_q)\overline{P}}{N^e+\sum\limits_{j\in\Nc_e}|h_{1,j}^e|^2\overline{P}})\\
&{=}& \log(\frac{N^e}{N^l}\frac{N^l+\|\mathbf{B}_q\|_2^2\|\mathbf{h}_q^\dag\mathbf{V}_q\|^2_2\overline{P}}{N^e+\sum\limits_{j\in\Nc_e}|h_{1,j}^e|^2\overline{P}})\\
&\stackrel[n_l\rightarrow\infty]{(b)}{\geq}&\log\frac{1}{n_e}(\frac{d_e}{d_c})^{\alpha}\stackrel{(c)}{\rightarrow}\infty\quad \mbox{as}\quad n_l\rightarrow\infty\yesnumber\label{eqn:SL_ser_2_c}
\end{IEEEeqnarray*}
(a) is obtained by defining $\mathbf{h}_q=[h_{1,{i+1}}^l,\ldots,h_{q,{i+1}}^l]^T$. (b) follows from $\|\mathbf{B}_q\|_2^2\geq \mathbf{B}_q(q)=\beta'_{qq}=1$, $\|\mathbf{h}_q\|_2^2\geq |h_{q,{i+1}}^l|^2\geq  d_c^{-\alpha}$, $\|\mathbf{V}_q\|_2^2=1$ and the randomness of $\mathbf{H}_{\Nc_e,\Tc^q}$ and $\mathbf{h}_q$. (c) is due to \eqref{eqn:ser_squ-c}. This completes the proof.
\end{IEEEproof}

\section{Discussion and Conclusion}\label{sec:discussion}
\textbf{Comparison to existing results}:
In general, we can define the cost of secure communication as $\mathfrak{C}_s=\frac{\overline{P}_{tot}}{\Rc_{s}}$. In prior works \cite{KoyKokElg12,ZhaFuWan12,VasGoeTow10,CapGoeLiuTow12,SheGoePisTow12}, due to the individual power constraint (the transmission power for each node is fixed), $\overline{P}_{tot}$ scales linearly with the number of nodes. Therefore, the scaling for the cost of secure communication lies in $[\sqrt{n},n]$ and $\mathfrak{C}_s\rightarrow\infty$ as $n\rightarrow\infty$. Here, we showed that cooperation based schemes can achieve secure communication with cost that goes to 0 as the number of nodes goes to $\infty$. This is so because we use a fixed $\overline{P}_{tot}$. Table~\ref{tbl:comparison} compares our scaling result to the existing ones, for both colluding and non-colluding eavesdroppers. It can be seen that in both interference-limited and cooperative network models, the same secure communication cost can be achieved, by tolerating a slightly lower number of eavesdroppers for the colluding case (compared to the non-colluding case). However, in the cooperative model, this degradation depends on the path loss exponent, $\alpha>2$, and it improves as $\alpha$ increases. Moreover, the scaling of  Theorem~\ref{thm:SL_ser} can be written as $\frac{n_e}{n_l}=o((n_e(\log n_e)^{\gamma})^{-1})\simeq o(n_e^{-1})$. To compare the scalings, one must compare $(\log{n_l})^{2}$ (for result of \cite{KoyKokElg12}) with $n_e$ (for our result), which seem to have the same order. For example, $n_e=\log{n_l}$ satisfies both scalings.

Zero cost communication \emph{with no secrecy constraint} was achieved in \cite{XieKum04}. Note that our strategy tolerates $n_e$ eavesdroppers satisfying the constraints in Table~\ref{tbl:comparison}. Hence, compared to \cite{XieKum04}, this means that this number of eavesdroppers does not affect the scaling of communication cost.

\begin{table*}
\renewcommand{\arraystretch}{1.3}
\caption{}
\label{tbl:comparison} \centering
\begin{tabular}{|c|c|c|}
\hline
& Non-colluding   &  Colluding\\
\hline
Interference-limited \cite{KoyKokElg12}, $\mathfrak{C}_s\rightarrow\infty$& $\frac{n_e}{n_l}=o((\log{n_l})^{-2})$  & $\frac{n_e}{n_l}=O((\log{n_l})^{-2-\rho}),\rho>0$\\
\hline
Cooperative, $\mathfrak{C}_s\rightarrow 0$& Theorem~\ref{thm:SL_ser}: $n_e^2(\log n_e)^{\gamma}=o(n_l)$ & Theorem~\ref{thm:SL_ser-c}: $n_e^{(2+\frac{2}{\alpha})}(\log n_e)^{\gamma}=o(n_l),\gamma>0$\\
\hline
\end{tabular}
\end{table*}

\textbf{Parallel vs. serial relaying:} Apart from the difference in the derived scaling for the number of tolerated eavesdroppers, our two schemes differ in terms of the individual power allocation. The parallel relaying scheme uses fewer relay nodes than the serial scheme. Hence, a larger fraction of $\overline{P}_{tot}$ is allocated per node. Therefore, serial relaying may be suitable for power-limited applications, with strict per node power constraints. For both schemes, the per node allocated power vanishes as the number of nodes increases but with different asymptotic behavior.

\textbf{Channel State Information (CSI):} In our network model ($\Sc\Nc$ and $\Sc\Nc$-PCE, notably \eqref{eqn:chgain}), CSI is equivalent to node location information. CSI for legitimate nodes can be obtained in practice (e.g., pilot symbols, feedback). The challenge is to obtain the eavesdroppers' CSI.
We assume global CSI is available, a common assumption in most of the physical layer security schemes (e.g., \cite{BasUlu12,BasUlu13}). DF relaying can be applied without the eavesdroppers' CSI as it only needs the location of the closest eavesdropper. However, to design the beamforming coefficients for ZF, full CSI is necessary.
Due to the complexity of the problem, this idealistic assumption allows to gain valuable insights. Obviously, the next step we consider as a future work is to investigate the problem when less or no eavesdroppers' CSI is available. Less CSI means knowledge of the eavesdroppers' channel statistics or imperfect estimates. In practice, these assumptions are appropriate in some scenarios, e.g., public safety, where some areas are less likely to have eavesdroppers. For imperfect CSI estimation, the authors in \cite{DonHanPetPoo08} showed that the achievable secrecy rate depends on the estimation error covariance matrix. Moreover, \cite{BloLan09} concludes that to achieve secure rate in wireless networks one needs little CSI.
We contrast our result of achieving infinite rate with known eavesdropper CSI/location, to the results for the \emph{interference-limited} channel model: if the location of eavesdroppers is unknown \cite{VasGoeTow10,CapGoeLiuTow12,SheGoePisTow12}, the achievable rate is of order~$1$.

\textbf{Perfect versus constrained collusion:}
We assumed \emph{perfect} colluding eavesdroppers, considering that the eavesdroppers share their observations freely. Collusion in \emph{large} wireless networks in all prior works is also assumed to be perfect \cite{KoyKokElg12,ZhaFuWan12}.
Recently, investigating the ramifications of the collusion models, the \emph{Wiretap Channel with Constrained Colluding Eavesdroppers} (WTC-CCE) was proposed \cite{MirPapIzs14}: two colluding eavesdroppers communicate over a virtual collusion channel, in addition to the main point-to-point communication channel (one legitimate transmitter-receiver pair). Extending the WTC-CCE to the model at hand can be a natural future work item; however, this is not trivial due to the complexity of both models.

\appendices
\section{Proof of Lemma~\ref{lemma:Ach_par_DF_DM}}\label{app:proof_lemma:Ach_par_DF_DM}
\begin{IEEEproof}
The proof is based on the random coding scheme, which combines Csiszar and Korner's scheme \cite{CsiKor78} and DF strategy (two-stage block Markov superposition coding). For decoding at the receivers we utilize backward decoding \cite{ElgKim11}.\footnote{This scheme first proposed for a relay-eavesdropper channel in \cite{LaiElG08} and extended to multiple relays with an eavesdropper in \cite{BasUlu12}.} We prove the theorem by using the replacement $X_1$ for $U_1$ and then the general case can be proved using a memoryless prefix channel as \cite{LaiElG08}.
Also, as mentioned in Remark~\ref{remark:agg_rate}, we assume $\Rc_{s}=R_1$. Consider a block Markov encoding scheme with $B$ blocks of transmission, each of $n_t$ symbols. A sequence of $B - 1$ messages, $m_{1,b}, b = 1, 2, \ldots,B - 1$, each selected independently and uniformly over $\Mc_1$ is to be sent over the channel in $n_tB$ transmissions. Note that as $B\rightarrow \infty$, the average rate $R_1(B-1)/B$ is arbitrarily close to $R_1$.

\textit{Codebook Generation:} Fix a joint p.m.f as \eqref{eqn:Ach_par_DF_DM_pmf}. Let $R_w=\min\{\min\limits_{i\in\Tc\setminus \{1\}} I(X_1;Y_{i}^l|U), I(U,X_1;Y_{n_{l}}^l)\}$. Generate $2^{n_tR_w}$ i.i.d $u^{n_t}$ sequences, each with probability $\prod\limits_{j = 1}^{n_t}p(u_{j})$. Index them as $u^{n_t}(m'_1,s')$ where $m'_1\in[1:2^{n_tR_1}]$ and $s'\in[1:2^{n_t(R_w-R_1)}]$. For each $u^{n_t}(m'_1,s')$, generate $2^{n_tR_w}$ conditionally i.i.d $x_1^{n_t}$ sequences, according to  probability $\prod\limits_{j = 1}^{n_t}p(x_{1,j}|u_{j})$. Index them as $x_1^{n_t}(m'_1,s',m_1,s)$, where $m_1\in[1:2^{n_tR_1}]$ and $s\in[1:2^{n_t(R_w-R_1)}]$. $s$ is the randomness index used to protect $u$ based on the Wyner code partitioning method.

\textit{Encoding (at the beginning of block $b$):} Let $m_{1,b}$ be the new message to be sent from the source node in block $b$. The stochastic encoder at the source uniformly randomly chooses $s_b$ and transmits $x_1^{n_t}(m_{1,b-1},s_{b-1},m_{1,b},s_b)$. Each relay node $i\in\Tc\setminus \{1\}$ knows the estimates $m_{1,b-1},s_{b-1}$ of the messages the source sent in the previous block; hence, it picks $u^{n_t}(m_{1,b-1},s_{b-1})$ and sends $\prod\limits_{j=1}^{n_t}p(x_{i,j}|u_{j})$. We assume that in the first block, cooperative information is $m_{1,b-1}=m_{1,0}=1$ and in the last block, a previously known message $m_{1,b}=m_{1,B}=1$ is transmitted.

\textit{Decoding (at the end of block $b$):} Each relay node $i\in\Tc\setminus \{1\}$ wants to correctly recover $m_{1,b},s_{b}$. Hence, it seeks a unique pair $(\tilde{m}_{1,b},\tilde{s}_{b})$ such that
\begin{IEEEeqnarray*}{c}
(x_1^{n_t}(m_{1,b-1},s_{b-1},\tilde{m}_{1,b},\tilde{s}_b),u^{n_t}(m_{1,b-1},s_{b-1}),y_i^l(b))\in A_\epsilon^{n_t}(X_1,U,Y_i^l)
\end{IEEEeqnarray*}
Considering the first term of $R_w$ and using the covering lemma \cite{ElgKim11}, this can be done with small enough probability of error if $n_t$ is sufficiently large. Backward decoding is used at the destination node $n_l$, hence it starts decoding after all $B$ blocks are received. Using its channel output at the end of block $b$, i.e., $y_{n_l}^l(b)$, the decoder at the destination looks for a unique pair $\tilde{m}_{1,b-1},\tilde{s}_{b-1}$ and such that
\begin{IEEEeqnarray*}{c}
(x_1^{n_t}(\tilde{m}_{1,b-1},\tilde{s}_{b-1},m_{1,b},s_b),u^{n_t}(\tilde{m}_{1,b-1},\tilde{s}_{b-1}),y_{n_l}^l(b))\in A_\epsilon^{n_t}(X_1,U,Y_{n_l}^l)
\end{IEEEeqnarray*}
where $m_{1,b},s_b$ were decoded in the previous step of backward decoding (i.e., block $b+1$). Similarly, considering the second term in $R_w$ for large enough $n_t$, the probability of error can be made sufficiently small.

\textit{Analysis of information leakage rate:} Let $j^*=\argmin\limits_{j\in\Nc_e}\{R_w-I(U,U_1;Y_{j}^e)\}$ and consider the mutual information between $M_1$ and $(Y_{j^*}^e)^{n_t}\doteq \mathbf{Y}^e$, averaged over the random codebook $\Cc$.
\begin{IEEEeqnarray*}{rcl}
I(M_1;\mathbf{Y}^e|\Cc)&=&H(M_1|\Cc)-H(M_1|\mathbf{Y}^e,\Cc)\\
&=&n_tR_1-H(M_1,S|\mathbf{Y}^e,\Cc)+H(S|M_1,\mathbf{Y}^e,\Cc)\\
&=&n_tR_1-H(M_1,S|\Cc)+I(M_1,S;\mathbf{Y}^e|\Cc)+H(S|M_1,\mathbf{Y}^e,\Cc)\\
&=&n_tR_1-n_tR_w+I(M_1,S,U^{n_t},X_1^{n_t};\mathbf{Y}^e|\Cc)+H(S|M_1,\mathbf{Y}^e,\Cc)\\
&\leq& n_tR_1-n_tR_w+I(M_1,S,U^{n_t},X_1^{n_t},\Cc;\mathbf{Y}^e)+H(S|M_1,\mathbf{Y}^e,\Cc)\\
&\stackrel{(a)}{\leq}& n_tR_1-n_tR_w+n_tI(U,X_1;Y_{j^*}^e)+H(S|M_1,\mathbf{Y}^e,\Cc)\\
&\stackrel{(b)}{\leq}& n_t(R_1-R_w+I(U,X_1;Y_{j^*}^e)+R_w-R_1-I(U,X_1;Y_{j^*}^e)+\varepsilon)\\
&\leq& n_t\varepsilon
\end{IEEEeqnarray*}
(a) holds because $M_1,S,\Cc\rightarrow U^{n_t},X_1^{n_t}\rightarrow\mathbf{Y}^e$ forms a Markov chain and thanks to the memoryless property. (b) follows because by using \cite[Lemma~22.1]{ElgKim11}, we have: if $R_w-R_1\geq I(U,X_1;Y_{j^*}^e)$, then $H(S|M_1,\mathbf{Y}^e,\Cc)\leq n_t(R_w-R_1-I(U,X_1;Y_{j^*}^e)+\varepsilon)$. This condition, after applying prefix channel, gives \eqref{eqn:Ach_par_DF_DM}.
\end{IEEEproof}

\section{Proof of Lemma~\ref{lemma:Ach_par_DF}}\label{app:proof_lemma:Ach_par_DF}
\begin{IEEEproof}
The achievable secrecy rate in Lemma~\ref{lemma:Ach_par_DF_DM} can be extended to the Gaussian case with continuous alphabets (and so to our network model) by standard arguments \cite{CovTho06}. We constrain all the inputs to be Gaussian.
For certain  $\beta_i, i\in[1:n_r+1]$, consider the following mapping for the generated codebook in Lemma~\ref{lemma:Ach_par_DF_DM} with respect to the p.m.f \eqref{eqn:Ach_par_DF_DM_pmf}, which contains a simple Gaussian version of the block Markov superposition coding where all relay nodes send the same common RV (shown by $U$). However, they adjust their power and use beamforming.
\begin{IEEEeqnarray*}{lcl}
U\sim\Cc\Nc(0,\tilde{P}_u)&\quad\mbox{and}\quad&\tilde{U}_1\sim\Cc\Nc(0,\tilde{P}_1)\\
X_1=\tilde{U}_1+\beta_1U&\mbox{and}&X_i=\beta_iU,i\in[2:n_r+1]
\end{IEEEeqnarray*}
Parameter $\beta_1$ determines the amount of $\tilde{P}_1$ dedicated to construct the basis of cooperation, while parameters $\beta_i,i\in[2:n_r+1]$ are the beamforming coefficients. Applying the power constraint in (\ref{eqn:tot_power_cons}) to above mapping, we obtain
\begin{IEEEeqnarray}{rcl}
\tilde{P}_1+\|\mathbf{B}\|_2^2\tilde{P}_u\leq \overline{P}_{tot}.\label{eqn:tot_power_cons_Ach_par_DF}
\end{IEEEeqnarray}
Now, it is sufficient to evaluate the mutual information terms in \eqref{eqn:Ach_par_DF_DM} by using this mapping and the network model in \eqref{eqn:RxLeg} and \eqref{eqn:RxEav}, to reach \eqref{eqn:Ach_par_DF}.
\end{IEEEproof}

\section{}\label{app:lemma:Poisson_che}
\begin{lemma}\label{lemma:Poisson_che}
Consider a Poisson RV $X$ with parameter $\lambda$.
\begin{IEEEeqnarray*}{c}
Pr(X\geq x)\leq \frac{e^{-\lambda}(e\lambda)^x}{x^x}\quad \mbox{for}\quad x>\lambda
\end{IEEEeqnarray*}
And hence
For any $\epsilon\in(0,1)$:
\begin{IEEEeqnarray}{c}
\lim\limits_{\lambda\rightarrow\infty}Pr(X\leq (1-\epsilon)\lambda)=0,\label{eqn:Poisson_che-}\\
\lim\limits_{\lambda\rightarrow\infty}Pr(X\leq (1+\epsilon)\lambda)=1.\label{eqn:Poisson_che+}
\end{IEEEeqnarray}
\end{lemma}
\begin{IEEEproof}
See \cite{KoyKokElg12} for proofs based on applying Chernoff bound and Chebyshev's inequality.
\end{IEEEproof}

\section{Proof of Lemma~\ref{lemma:Ach_ser_DF_DM}}\label{app:proof_lemma:Ach_ser_DF_DM}
\begin{IEEEproof}
The proof is similar to the proof of Lemma~\ref{lemma:Ach_par_DF_DM}. Therefore, we only highlight the differences. Here, a serial $(n_l-1)$-stage block Markov coding is used.\footnote{The scheme for the multiple relay networks is first proposed in \cite{XieKum04,XieKum05} and then extended to multiple relays with an eavesdropper in \cite{BasUlu12}.} Without loss of generality and for simplicity, we choose the identity permutation and prove the achievability of
\begin{IEEEeqnarray*}{l}
\min\limits_{j\in\Nc_e}\min\limits_{i\in[1:n_l-1]}I(U_1^i;Y_{i+1}^l|U_{i+1}^{n_l-1})-I(U_1^{n_l-1};Y_{j}^e)
\end{IEEEeqnarray*}
Moreover, we use the replacement $X_i$ for $U_i$ in the proof and then the general case can be proved using a memoryless prefix channel as \cite{LaiElG08}.                                  A sequence of $B - n_l+2$ messages, $m_{1,b}, b = 1, 2, \ldots,B - n_l+2$, each selected independently and uniformly over $\Mc_1$ is to be sent over the channel in $n_tB$ transmission. Note that as $B\rightarrow \infty$, the average rate $R_1(B- n_l+2)/B$ is arbitrarily close to $R_1$.

\textit{Codebook Generation:} Fix a joint p.m.f as \eqref{eqn:Ach_ser_DF_DM_pmf}. Define $w_\iota=(m_{1,\iota},s_{\iota})$ where $m_{1,\iota}\in[1:2^{n_tR_1}]$ and $s_{\iota}\in[1:2^{n_t(R_w-R_1)}]$ and $R_w=\min\limits_{i\in[1:n_l-1]} I(X_1^i;Y_{i+1}^l|X_{i+1}^{n_l-1})$. Generate $2^{n_tR_w}$ i.i.d $x_{n_l-1}^{n_t}(w_{n_l-1})$
where $w_{n_l-1}\in[1:2^{n_tR_w}]$. For each $x_{n_l-1}^{n_t}(w_{n_l-1})$, generate $2^{n_tR_w}$ conditionally i.i.d $x_{n_l-2}^{n_t}(w_{n_l-1},w_{n_l-2})$
where $w_{n_l-2}\in[1:2^{n_tR_w}]$. Continuing in this way, at each node $i\in[1:n_l-2]$, for each $(x_{i+1}^{n_t}(w_{i+1},\ldots$ $,w_{n_l-1}),\ldots,x_{n_l-1}^{n_t}(w_{n_l-1}))$ generate $2^{n_tR_w}$ conditionally i.i.d $x_{i}^{n_t}(w_{i},\ldots,w_{n_l-1})$ where $w_{i}\in[1:2^{n_tR_w}]$. Since we use sliding-window decoding, we repeat this process $n_l-1$ times and generate $n_l-1$ random codebooks which in block $b$ we use the $(b\mod n_l-1)$th codebook to make the error events independent.

\textit{Encoding (at the beginning of block $b$):} Let $m_{1,b}$ be the new message to be sent from the source node in block $b$. The stochastic encoder at the source uniformly randomly chooses $s_b$ and setting $w(b)=(m_{1,b},s_b)$ transmits $x_1^{n_t}(w(1),\ldots,w(n_l-1))$. Each node $i\in[1:n_l-1]$ knows the estimations of $w(b-r+1),r\geq i+1$ (from the decoding part) and transmits $x_{i}^{n_t}(w(b-i),\ldots,w(b-n_l+1))$. We assume that $w(b)=1, B-n_l+3\leq b\leq B$.

\textit{Decoding (at the end of block $b$):} Each node $i\in[2:n_l]$ wants to correctly recover $w(b-i+1)$. Hence, it looks for a unique index $\tilde{w}(b-i+1)$ such that for all $k=1,\ldots,i-1$ satisfy
\begin{IEEEeqnarray*}{c}
(x_{i-1-k}^{n_t}(\tilde{w}(b-i+1),w(b-i),\ldots,w(b-k-n_l+1)),\ldots,x_{n_l-1}^{n_t}(w(b-k-n_l+1)),y_{i}^l(b-k))\\
\in A_\epsilon^{n_t}(X_{i-1-k},\ldots,X_{n_l-1},Y_i^l)
\end{IEEEeqnarray*}
If $n_t$ is large enough, it can be shown from $R_w$, the covering lemma \cite{ElgKim11} and the independent codebooks over $(n_l-1)$ adjacent block, the probability of error can be made sufficiently small. The analysis of the information leakage rate can be done same as in the proof of Lemma~\ref{lemma:Ach_par_DF_DM}, by defining $U={X_2^{n_l-1}}$.
\end{IEEEproof}

\section{Proof of Lemma~\ref{lemma:Ach_ser_DF}}\label{app:proof_lemma:Ach_ser_DF}
\begin{IEEEproof}
Similar to the proof of Lemma~\ref{lemma:Ach_par_DF}, we compute \eqref{eqn:Ach_ser_DF_DM}, with an appropriate choice of the input distribution by constraining all the inputs to be Gaussian. For each $q\in[1:n_l-1]$, define $\mathbf{B}_q=[\beta'_{1q},\ldots,\beta'_{qq}]\in\mathbb{C}^{q}$ for $\beta'_{qq}=1$ and certain $\beta'_{kq}, k\in[1:q-1]$ and consider the following mapping for the generated codebook in Lemma~\ref{lemma:Ach_ser_DF_DM} with respect to the p.m.f \eqref{eqn:Ach_ser_DF_DM_pmf},
\begin{IEEEeqnarray}{lcl}
\tilde{U}_q\sim\Cc\Nc(0,\tilde{P}_q)&\:,\:&q\in[1:n_l-1]\label{eqn:map1_Ach_ser_DF}\\
X_k=\sum\limits_{q=k}^{n_l-1}\beta'_{kq}\tilde{U}_q=\tilde{U}_k+\sum\limits_{q=k+1}^{n_l-1}\beta'_{kq}\tilde{U}_q&,&k\in[1:n_l-1]\label{eqn:map2_Ach_ser_DF}
\end{IEEEeqnarray}
Each node $k$ (considering the ordered set of transmitters $k\in[1:n_l-1]$) in each block $b$ transmits a linear combination of the decoded codewords in the $n_l-k$ previous blocks (shown by $\tilde{U}_q(w_{b-q+1}),k\leq q\leq n_l-1$). These codewords make the coherent transmission between this node $k$ and node $i,1\leq i<k$ to each node $q, k<q\leq n_l-1$. Beamforming using parameters $\beta'_{kq}$ is applied by adjusting the power of these codewords. Applying the power constraint in (\ref{eqn:tot_power_cons}) to the above mapping, we obtain
\begin{IEEEeqnarray*}{rcl}
\overline{P}_{tot}&\geq&\sum\limits_{k=1}^{n_l-1}\sum\limits_{q=k}^{n_l-1}|\beta'_{kq}|^2\tilde{P}_q=\sum\limits_{q=1}^{n_l-1}\sum\limits_{k=1}^{q}|\beta'_{kq}|^2\tilde{P}_q=\sum\limits_{q=1}^{n_l-1}\|\mathbf{B}_q\|_2^2\tilde{P}_q
\yesnumber\label{eqn:tot_power_cons_Ach_ser_DF}
\end{IEEEeqnarray*}
Using this mapping, \eqref{eqn:RxLeg} and \eqref{eqn:RxEav}, and applying interchangings in the order of summations similar to \eqref{eqn:tot_power_cons_Ach_ser_DF}, deriving the mutual information terms in \eqref{eqn:Ach_ser_DF_DM} completes the proof.
\end{IEEEproof}

\section{Proof of Theorem~\ref{thm:Ach_ser_DF-c}}\label{app:proof_thm:Ach_ser_DF-c}
\begin{IEEEproof}
First, we consider a Discrete Memoryless version of the $\Sc\Nc$-PCE and derive an achievable secure aggregate rate $\Rc_s^{DM}$ in Lemma~\ref{lemma:Ach_ser_DF_DM-c}. The proof is based on using $(n_l-1)$-stage block Markov coding (serial DF relaying) and Wyner wiretap coding and is given in Appendix~\ref{app:proof_lemma:Ach_ser_DF_DM-c}. Without loss of generality, let $\Nc_l=\{1,\ldots,n_l\}$. In the serial relaying scheme, the transmitted signal of each node $i$ can be decoded in all subsequent nodes ($i+1$ to $n_l$). Hence, it can decode the transmitted signals of nodes 1 to $i-1$ \cite{KraGasGup05}.
Next, we extend $\Rc_s^{DM}$ to $\Sc\Nc$-PCE in Lemma~\ref{lemma:Ach_ser_DF-c} and call it $\Rc_s$. Finally, we apply ZF to $\Rc_s$ and we obtain $\Rc_s^{ZF}$.
Similar to the non-colluding case, we need clustering to apply ZF at the eavesdroppers. Each cluster determines the priority of decoding (starting from the source node).
This means that the nodes in each cluster form a group of relays with the same priority; this enables them to act as a distributed multi-antenna to collectively apply ZF.
Here, our rate expressions show the collusion effect. In Step~2, we adjust the size of the clusters to combat the collusion effect.

\begin{lemma}\label{lemma:Ach_ser_DF_DM-c}
Consider the general discrete memoryless counterpart of $\Sc\Nc$-PCE, given by some conditional distribution $p(y_2^l,\ldots,y_{n_l}^l,\mathbf{y}^e|x_1,\ldots,x_{n_l})$, and let $\pi(\cdot)$ be a permutation on $\Nc_l=\{1,\ldots,n_l\}$, where $\pi(1)=1$, $\pi(n_l)=n_l$ and $\pi(m:n)=\{\pi(m),\pi(m+1),\ldots,\pi(n)\}$. The secrecy capacity is lower-bounded by:
\begin{IEEEeqnarray}{rl}\label{eqn:Ach_ser_DF_DM-c}
\Rc_s^{DM}=\sup\max\limits_{\pi(\cdot)}\min\limits_{i\in[1:n_l-1]}&I(U_{\pi(1:i)};Y_{\pi(i+1)}^l|U_{\pi(i+1:n_l-1)})-I(U_{\pi(1:n_l-1)};\mathbf{Y}^e)
\end{IEEEeqnarray}
where the supremum is taken over all joint p.m.fs of the form
\begin{IEEEeqnarray}{c}\label{eqn:Ach_ser_DF_DM_pmf-c}
p(u_1,\ldots,u_{n_l-1})\prod\limits_{k=1}^{n_l-1} p(x_k|u_k).
\end{IEEEeqnarray}
\end{lemma}
Now, we extend the above lemma to $\Sc\Nc$-PCE, using an appropriate codebook mapping based on Gaussian RVs in the following lemma (proof is provided in Appendix~\ref{app:proof_lemma:Ach_ser_DF-c}).
\begin{lemma}\label{lemma:Ach_ser_DF-c}
For $\Sc\Nc$-PCE, the following secure aggregate rate is achievable.
\begin{IEEEeqnarray*}{rl}
\Rc_s=\min\limits_{i\in[1:n_l-1]}\max_{\mathbf{B}_i,\tilde{P}_i}&\log(1+\frac{\sum\limits_{q=1}^i|\sum\limits_{k=1}^qh_{k,{i+1}}^l\beta'_{kq}|^2\tilde{P}_q}{N^l})
-\log(1+\frac{\sum\limits_{j\in\Nc_e}\sum\limits_{q=1}^{n_l-1}|\sum\limits_{k=1}^qh_{k,{j}}^e\beta'_{kq}|^2\tilde{P}_q}{N^e})\yesnumber\label{eqn:Ach_ser_DF-c}
\end{IEEEeqnarray*}
where \eqref{eqn:ach_coef_ser-c} and \eqref{eqn:ach_ZF_cond_pow_ser-c} hold.
\end{lemma}
The serial relaying scheme overcomes the decoding constraint at the farthest relay by ordering the relays. Hence, all nodes in the network (except the source and destination) can be used as the relays; thus, $\Tc=\{1,\ldots,n_l-1\}$. From \eqref{eqn:Ach_ser_DF-c}, we see that the optimal beamforming strategy is the one that results in $\max$ over the beamforming coefficient vector $\mathbf{B}$. Finding the closed form solution is an open problem \cite{DonHanPetPoo08}. Hence, we choose to ZF at the colluding eavesdroppers by letting $\sum\limits_{q=2}^{n_l-1}|\sum\limits_{k=1}^qh_{k,{j}}^e\beta'_{kq}|^2\tilde{P}_q=0,\forall j\in\Nc_e$. This results in $\tilde{P}_q=0$ or $E(q,j)=\sum\limits_{k=1}^qh_{k,{j}}^e\beta'_{kq}=0$,
for $\forall q\in[2:n_l-1]$.
Now we show that indeed clustering is needed by deriving the power allocation in \eqref{eqn:ach_ZF_cond_cluspow_ser-c}. One can obtain $X_k=\tilde{U}_k+\beta_{k}X_{k+1}$ from \eqref{eqn:map2_Ach_ser_DF-c}, where $\beta'_{kq}=\prod\limits_{m=k}^{q-1}\beta_m$. Therefore, it is seen that $E(q_0,j)$ and $E(q_0+1,j)$ only differ in one variable, i.e., $\beta_{q_0+1}$.
However, to apply ZF, $E(q,j)$ must be equal to zero for all $j\in\Nc_e$ if $\tilde{P}_q>0$, which is clearly not possible. Therefore, we set $\tilde{P}_q=0$ if $q\bmod n_e\neq1$ and leave only one equation needing to be satisfied, i.e, $E(q,j)=0$ if $q\bmod n_e=1\forall j\in\Nc_e$, in every $n_e$ equations. Therefore, power allocation in \eqref{eqn:ach_ZF_cond_cluspow_ser-c} makes the ZF possible. Thus, the coefficient vector $\mathbf{B}_q$ must lie in the null space of $\mathbf{H}_{\Nc_e,\Tc^q}$, i.e, $\mathbf{H}_{\Nc_e,\Tc^q}\mathbf{B}_q=\mathbf{0}$, which is given in \eqref{eqn:ach_ZF_cond_coef_ser-c}. \eqref{eqn:Ach_ser_ZF-c} is resulted from applying \eqref{eqn:ach_ZF_cond_coef_ser-c} on \eqref{eqn:Ach_ser_DF-c}.
To summarize: in order to overcome $n_e$ eavesdroppers, every $n_e$ nodes form a cluster, where they transmit the same information in each block (equal part of fresh information) and they apply beamforming to ZF all eavesdroppers.
To complete the proof, it is enough to derive the total power constraint \eqref{eqn:ach_ZF_cond_pow_ser-c} already given in Lemma~\ref{lemma:Ach_ser_DF-c}.
\end{IEEEproof}

\section{Proof of Lemma~\ref{lemma:Ach_ser_DF_DM-c}}\label{app:proof_lemma:Ach_ser_DF_DM-c}
\begin{IEEEproof}
The proof is similar to Lemma~\ref{lemma:Ach_ser_DF_DM}. The difference is in the analysis of the information leakage rate. Therefore, we only provide this analysis for brevity. Consider the mutual information between $M_1$ and $(\mathbf{Y}^e)^{n_t}$, averaged over the random codebook $\Cc$.
\begin{IEEEeqnarray*}{rcl}
&I&(M_1;(\mathbf{Y}^e)^{n_t}|\Cc)=H(M_1|\Cc)-H(M_1|(\mathbf{Y}^e)^{n_t},\Cc)\\
&=&n_tR_1-H(M_1,S|(\mathbf{Y}^e)^{n_t},\Cc)+H(S|M_1,(\mathbf{Y}^e)^{n_t},\Cc)\\
&=&n_tR_1-H(M_1,S|\Cc)+I(M_1,S;(\mathbf{Y}^e)^{n_t}|\Cc)\\
&&+H(S|M_1,(\mathbf{Y}^e)^{n_t},\Cc)\\
&=&n_tR_1-n_tR_w+I(M_1,S,U^{n_t},X_1^{n_t};(\mathbf{Y}^e)^{n_t}|\Cc)\\
&&+H(S|M_1,(\mathbf{Y}^e)^{n_t},\Cc)\\
&\leq& n_tR_1-n_tR_w+I(M_1,S,U^{n_t},X_1^{n_t},\Cc;(\mathbf{Y}^e)^{n_t})\\
&&+H(S|M_1,(\mathbf{Y}^e)^{n_t},\Cc)\\
&\stackrel{(a)}{\leq}& n_tR_1-n_tR_w+n_tI(U,X_1;\mathbf{Y}^e)+H(S|M_1,(\mathbf{Y}^e)^{n_t},\Cc)\\
&\stackrel{(b)}{\leq}& n_t(R_1-R_w+I(U,X_1;\mathbf{Y}^e)+R_w-R_1-I(U,X_1;\mathbf{Y}^e)+\varepsilon)\\
&\leq& n_t\varepsilon
\end{IEEEeqnarray*}
(a) holds because $M_1,S,\Cc\rightarrow U^{n_t},X_1^{n_t}\rightarrow(\mathbf{Y}^e)^{n_t}$ forms a Markov chain and thanks to the memoryless property. (b) follows because by using \cite[Lemma~22.1]{ElgKim11}, we have: if $R_w-R_1\geq I(U,X_1;\mathbf{Y}^e)$, then $H(S|M_1,(\mathbf{Y}^e)^{n_t},\Cc)\leq n_t(R_w-R_1-I(U,X_1;\mathbf{Y}^e)+\varepsilon)$.
\end{IEEEproof}

\section{Proof of Lemma~\ref{lemma:Ach_ser_DF-c}}\label{app:proof_lemma:Ach_ser_DF-c}
\begin{IEEEproof}
Using standard arguments, we can extend \eqref{eqn:Ach_ser_DF_DM-c}, by computing it for an appropriate choice of the input distribution and constraining all the inputs to be Gaussian \cite{CovTho06}. The mapping is same as the one in Lemma~\ref{lemma:Ach_ser_DF}, which is repeated here for completeness (since it is needed in deriving the beamforming vector).

For each $q\in[1:n_l-1]$, define $\mathbf{B}_q=[\beta'_{1q},\ldots,\beta'_{qq}]\in\mathbb{C}^{q}$ for $\beta'_{qq}=1$ and certain $\beta'_{kq}, k\in[1:q-1]$ and consider the following mapping for the generated codebook in Lemma~\ref{lemma:Ach_ser_DF_DM-c} with respect to the p.m.f \eqref{eqn:Ach_ser_DF_DM_pmf-c},
\begin{IEEEeqnarray}{lcl}
\!\!\!\!\!\!\!\tilde{U}_q\sim\Cc\Nc(0,\tilde{P}_q)&\:,\:&q\in[1:n_l-1]\label{eqn:map1_Ach_ser_DF-c}\\
\!\!\!\!\!\!\!X_k=\sum\limits_{q=k}^{n_l-1}\beta'_{kq}\tilde{U}_q=\tilde{U}_k+\sum\limits_{q=k+1}^{n_l-1}\beta'_{kq}\tilde{U}_q&,&k\in[1:n_l-1]\label{eqn:map2_Ach_ser_DF-c}
\end{IEEEeqnarray}
Each node $k$ (considering the ordered set of transmitters $k\in[1:n_l-1]$) in each block $b$ transmits a linear combination of the decoded codewords in the $n_l-k$ previous blocks (shown by $\tilde{U}_q(w_{b-q+1}),k\leq q\leq n_l-1$). These codewords make the coherent transmission between this node $k$ and node $i,1\leq i<k$ to each node $q, k<q\leq n_l-1$. Beamforming using parameters $\beta'_{kq}$ is applied by adjusting the power of these codewords. Applying the power constraint in (\ref{eqn:tot_power_cons}) to the above mapping, we obtain
\begin{IEEEeqnarray*}{rcl}
\overline{P}_{tot}&\geq&\sum\limits_{k=1}^{n_l-1}\sum\limits_{q=k}^{n_l-1}|\beta'_{kq}|^2\tilde{P}_q=\sum\limits_{q=1}^{n_l-1}\sum\limits_{k=1}^{q}|\beta'_{kq}|^2\tilde{P}_q=\sum\limits_{q=1}^{n_l-1}\|\mathbf{B}_q\|_2^2\tilde{P}_q
\end{IEEEeqnarray*}
Using this mapping, \eqref{eqn:RxLeg} and \eqref{eqn:RxEav}, applying interchangings in the order of summations, and deriving the mutual information terms in \eqref{eqn:Ach_ser_DF_DM-c} completes the proof.
\end{IEEEproof}

\bibliographystyle{./IEEEtran}
\bibliography{./IEEEabrv,./SL_journal}

\begin{thebibliography}{10}
\providecommand{\url}[1]{#1}
\csname url@samestyle\endcsname
\providecommand{\newblock}{\relax}
\providecommand{\bibinfo}[2]{#2}
\providecommand{\BIBentrySTDinterwordspacing}{\spaceskip=0pt\relax}
\providecommand{\BIBentryALTinterwordstretchfactor}{4}
\providecommand{\BIBentryALTinterwordspacing}{\spaceskip=\fontdimen2\font plus
\BIBentryALTinterwordstretchfactor\fontdimen3\font minus
  \fontdimen4\font\relax}
\providecommand{\BIBforeignlanguage}[2]{{%
\expandafter\ifx\csname l@#1\endcsname\relax
\typeout{** WARNING: IEEEtran.bst: No hyphenation pattern has been}%
\typeout{** loaded for the language `#1'. Using the pattern for}%
\typeout{** the default language instead.}%
\else
\language=\csname l@#1\endcsname
\fi
#2}}
\providecommand{\BIBdecl}{\relax}
\BIBdecl

\bibitem{MirPapInf14}
M.~Mirmohseni and P.~Papadimitratos, ``Scaling laws for secrecy capacity in
  cooperative wireless networks,'' in \emph{Proc. IEEE INFOCOM}, Toronto,
  Canada, April 27 -- May 2, 2014.

\bibitem{MirPapIWCIT14}
------, ``Colluding eavesdroppers in large cooperative wireless networks,'' in
  \emph{Proc. Iran Workshop on Communication and Information Theory (IWCIT)},
  Tehran, Iran, May 2014.

\bibitem{ShiChaWuHuaChe11}
Y.~S. Shiu, S.~Y. Chang, H.~C. Wu, S.~C. Huang, and H.~H. Chen, ``Physical
  layer security in wireless networks: A tutorial,'' \emph{IEEE Wireless
  Communications}, vol.~18, no.~2, pp. 66--74, Apr. 2011.

\bibitem{ElgKim11}
A.~E. Gamal and Y.-H. Kim, \emph{Network information theory}.\hskip 1em plus
  0.5em minus 0.4em\relax Cambridge Univ. Press, 2011.

\bibitem{GupKum00}
P.~Gupta and P.~R. Kumar, ``The capacity of wireless networks,'' \emph{IEEE
  Trans. Inf. Theory}, vol.~46, no.~2, pp. 388--404, Mar. 2000.

\bibitem{FraDouTseThi07}
M.~Franceschetti, O.~Dousse, D.~N.~C. Tse, and P.~Thiran, ``Closing the gap in
  the capacity of wireless networks via percolation theory,'' \emph{IEEE Trans.
  Inf. Theory}, vol.~53, no.~3, pp. 1009--1018, Mar. 2007.

\bibitem{XieKum04}
L.-L. Xie and P.~R. Kumar, ``A network information theory for wireless
  communications: Scaling laws and optimal operation,'' \emph{IEEE Trans. Inf.
  Theory}, vol.~50, no.~5, pp. 748--767, May 2004.

\bibitem{OzgLevTse07}
A.~Ozgur, O.~Leveque, and D.~N.~C. Tse, ``Hierarchical cooperation achieves
  optimal capacity scaling in ad hoc networks,'' \emph{IEEE Trans. Inf.
  Theory}, vol.~53, no.~10, pp. 3549--3572, Oct. 2007.

\bibitem{KoyKokElg12}
O.~O. Koyluoglu, C.~E. Koksal, and H.~A.~E. Gamal, ``On secrecy capacity
  scaling in wireless networks,'' \emph{IEEE Trans. Inf. Theory}, vol.~58,
  no.~5, pp. 3000--3015, May 2012.

\bibitem{ZhaFuWan12}
J.~Zhang, L.~Pu, and X.~Wang, ``Impact of secrecy on capacity in large-scale
  wireless networks,'' in \emph{Proc. IEEE INFOCOM}, Orlando, FL, USA, Mar.
  2012, pp. 3051--3055.

\bibitem{VasGoeTow10}
S.~Vasudevan, D.~Goeckel, and D.~Towsley, ``Security-capacity trade-off in
  large wireless networks using keyless secrecy,'' in \emph{Proc. eleventh ACM
  international symposium on Mobile ad hoc networking and computing, MobiHoc
  '10}, Ill., USA, Sep. 2010, pp. 21--30.

\bibitem{CapGoeLiuTow12}
C.~Capar, D.~Goeckel, B.~Liu, and D.~Towsley, ``Secret communication in large
  wireless networks without eavesdropper location information,'' in \emph{Proc.
  IEEE INFOCOM}, Orlando, FL, USA, Mar. 2012, pp. 1152 -- 1160.

\bibitem{SheGoePisTow12}
A.~Sheikoleslami, D.~Goeckel, H.~Pishro-Nik, and D.~Towsley, ``Physical layer
  security from inter-session interference in large wireless networks,'' in
  \emph{Proc. IEEE INFOCOM}, Orlando, FL, USA, Mar. 2012, pp. 1179 -- 1187.

\bibitem{LTE10}
LTE-A, \emph{3rd Generation Partnership Project; Technical Specification Group
  Radio Access Network; Evolved Universal Terrestrial Radio Access
  (EUTRA)}.\hskip 1em plus 0.5em minus 0.4em\relax 3GPP TR 36.806 V9.0.0, 2010.

\bibitem{SawKisMorNisTan10}
M.~Sawahashi, Y.~Kishiyama, A.~Morimoto, D.~Nishikawa, and M.~Tanno,
  ``Coordinated multipoint transmission/reception techniques for lte-advanced,
  coordinated and distributed mimo,'' \emph{IEEE Wireless Communications},
  vol.~17, no.~3, pp. 26–--34, June 2010.

\bibitem{Sha49}
C.~E. Shannon, ``Communication theory of secrecy systems,'' \emph{Bell Syst.
  Tech. J.}, vol.~28, pp. 656--–715, 1949.

\bibitem{Wyn75}
A.~D. Wyner, ``{The Wire-tap Channel},'' \emph{Bell Systems Technical Journal},
  vol.~54, no.~8, pp. 1355--1387, Jan. 1975.

\bibitem{CsiKor78}
I.~Csiszar and J.~Korner, ``Broadcast channels with confidential messages,''
  \emph{IEEE Trans. Inf. Theory}, vol.~24, no.~3, pp. 339--348, May 1978.

\bibitem{LiuMarSpaYat08}
R.~Liu, I.~Maric, P.~Spasojevic, and R.~D. Yates, ``Discrete memoryless
  interference and broadcast channels with confidential messages: secrecy rate
  regions,'' \emph{IEEE Trans. Inf. Theory}, vol.~54, no.~6, pp. 2493--2507,
  Jun. 2008.

\bibitem{EkrUlu13}
E.~Ekrem and S.~Ulukus, ``Multi-receiver wiretap channel with public and
  confidential messages,'' \emph{IEEE Trans. Inf. Theory}, vol.~59, no.~4, pp.
  2165--2177, April 2013.

\bibitem{ChiaElG12}
Y.~K. Chia and A.~E. Gamal, ``Three-receiver broadcast channels with common and
  confidential messages,'' \emph{IEEE Trans. Inf. Theory}, vol.~58, no.~5, pp.
  2748--2765, May 2012.

\bibitem{Ooh07}
Y.~Oohama, ``Capacity theorems for relay channels with confidential messages,''
  in \emph{Proc. IEEE Int. Symp. Info. Theory (ISIT)}, Nice, France, Jun. 2007.

\bibitem{LaiElG08}
L.~Lai and H.~E. Gamal, ``The relay-eavesdropper channel: cooperation for
  secrecy,'' \emph{IEEE Trans. Inf. Theory}, vol.~54, no.~9, pp. 4005--4019,
  Sep. 2008.

\bibitem{LiaPooYin11}
Y.~Liang, H.~V. Poor, and L.~Ying, ``Secure communications over wireless
  broadcast networks: stability and utility maximization,'' \emph{IEEE Trans.
  Inf. Forensics and Security}, vol.~6, no.~3, pp. 682--692, Sep. 2011.

\bibitem{BasUlu12}
R.~Bassily and S.~Ulukus, ``Secure communication in multiple relay networks
  through decode-and-forward strategies,'' \emph{Journal of Communications and
  Networks, special issue on Physical Layer Security}, vol.~14, no.~4, pp.
  352--363, Aug. 2012.

\bibitem{BasUlu13}
------, ``Deaf cooperation and relay selection strategies for secure
  communication in multiple relay networks,'' \emph{IEEE Trans. Signal
  Processing}, vol.~61, no.~6, pp. 1544--1554, Mar. 2013.

\bibitem{CoveElg79}
T.~M. Cover and A.~E. Gamal, ``Capacity theorems for relay channels,''
  \emph{IEEE Trans. Inf. Theory}, vol.~25, no.~5, pp. 572--584, Sep. 1979.

\bibitem{DonHanPetPoo08}
L.~Dong, Z.~Han, A.~P. Petropulu, and H.~V. Poor, ``Secure wireless
  communications via cooperation,'' in \emph{Proc. Allerton Conf. Commun.,
  Control, Comput.}, Monticello, IL, USA, Sep. 2008.

\bibitem{NegGoe05}
R.~Negi and S.~Goel, ``Secret communication using artificial noise,'' in
  \emph{Proc. IEEE VTC}, Sep. 2005, pp. 1906--1910.

\bibitem{LanWor03}
J.~N. Laneman and G.~W. Wornell, ``Distributed space-time-coded protocols for
  exploiting cooperative diversity in wireless networks,'' \emph{IEEE Trans.
  Inf. Theory}, vol.~49, no.~10, pp. 2415--2425, Oct. 2003.

\bibitem{shaLiuUlu09}
S.~Shafiee, N.~Liu, and S.~Ulukus, ``Towards the secrecy capacity of the
  gaussian {MIMO} wire-tap channel: The 2-2-1 channel,'' \emph{IEEE Trans. Inf.
  Theory}, vol.~55, no.~9, pp. 4033--–4039, Sep. 2009.

\bibitem{KhiWor10}
A.~Khisti and G.~Wornell, ``Secure transmission with multiple antennas i: The
  {MISOME} wiretap channel,'' \emph{IEEE Trans. Inf. Theory}, vol.~56, no.~7,
  pp. 3088--–3104, July 2010.

\bibitem{YukErk11}
M.~Yuksel and E.~Erkip, ``Diversity-multiplexing trade off for the
  multiple-antenna wire-tap channel,'' \emph{IEEE Trans. Wireless Commun.},
  vol.~10, no.~3, pp. 762–--771, Mar. 2011.

\bibitem{RezAlo12}
Z.~Rezki and M.~S. Alouini, ``Secure diversity-multiplexing tradeoff of
  zero-forcing transmit scheme at finite-snr,'' \emph{IEEE Trans. Commun.},
  vol.~60, no.~4, pp. 1138--–1147, Apr. 2012.

\bibitem{OggHas11}
F.~Oggier and B.~Hassibi, ``The secrecy capacity of the {MIMO} wiretap
  channel,'' \emph{IEEE Trans. Inf. Theory}, vol.~57, no.~8, pp. 4961–--4972,
  Aug. 2011.

\bibitem{PinBarWin09}
P.~C. Pinto, J.~Barros, and M.~Z. Win, ``Wireless physical-layer security: the
  case of colluding eavesdroppers,'' in \emph{Proc. IEEE Int. Symp. Info.
  Theory (ISIT)}, Seoul, Korea, Jun. 2009, pp. 2442--2446.

\bibitem{PinBarWin12II}
------, ``Secure communication in stochastic wireless networks part ii: maximum
  rate and collusion,'' \emph{IEEE Trans. Inf. Forensics and Security}, vol.~7,
  no.~1, pp. 139--147, Feb 2012.

\bibitem{XieKum05}
P.~R.~K. L.-L.~Xie, ``An achievable rate for the multiple-level relay
  channel,'' \emph{IEEE Trans. Inf. Theory}, vol.~51, no.~4, pp. 1348–--1358,
  Apr. 2005.

\bibitem{KraGasGup05}
G.~Kramer, M.~Gastpar, and P.~Gupta, ``Cooperative strategies and capacity
  theorems for relay networks,'' \emph{IEEE Trans. Inf. Theory}, vol.~51,
  no.~9, pp. 3037--3063, Sep. 2005.

\bibitem{WebAndJin10}
J.~A. S.~Weber and N.~Jindal, ``An overview of the transmission capacity of
  wireless networks,'' \emph{IEEE Trans. Commun.}, vol.~58, no.~12, pp.
  3593--–3604, Dec. 2010.

\bibitem{BloLan09}
M.~Bloch and J.~N. Laneman, ``Information-spectrum methods for
  information-theoretic security,'' in \emph{Proc. Inf. Theory and App.
  Workshop}, San Diego, CA, USA, Feb. 2009, pp. 23–--28.

\bibitem{MirPapIzs14}
M.~Mirmohseni and P.~Papadimitratos, ``Constrained colluding eavesdroppers: an
  information-theoretic model,'' in \emph{Proc. International Zurich Seminar on
  Communications (IZS)}, Zurich, Switzerland, Feb. 2014.

\bibitem{CovTho06}
T.~M. Cover and J.~A. Thomas, \emph{Elements of Information Theory}.\hskip 1em
  plus 0.5em minus 0.4em\relax Wiley-Interscience, 2006.

\end{thebibliography}

\end{document}